# Linear Programming in the Semi-streaming Model with Application to the Maximum Matching Problem

Kook Jin Ahn and Sudipto Guha*

May 29, 2018

## Abstract

In this paper we study linear programming based approaches to the maximum matching problem in the semi-streaming model. The semi-streaming model has been considered as one of the models for efficient processing massive graphs. In this model edges are presented sequentially, possibly in an adversarial order, and we are only allowed to use a small space. The allowed space is near linear in the number of vertices (and sublinear in the number of edges) of the input graph.

In recent years, there have been several new and exciting results in the semi-streaming model. However broad techniques such as linear programming have not been adapted to this model. In this paper we present several techniques to adapt and optimize linear programming based approaches in the semi-streaming model. We use the maximum matching problem as a foil to demonstrate the effectiveness of adapting such tools in this model and as a consequence we improve almost all previous results on the semi-streaming maximum matching problem. We also prove new results on interesting variants.

## 1 Introduction

Analyzing massive data sets has been one of the key motivations for studying streaming algorithms. In the streaming model we have sequential access to the input data and the random accessible memory is sublinear in the input size. In recent years, there has been significant progress in analyzing distributions in a streaming setting (see for example [25]), but similar progress has been elusive in the context of processing graph data. Massive graphs arise naturally in many disparate domains, for example, information retrieval, traffic and billing records of large networks, large scale scientific experiments, to name only a few. To help process such large graphs efficiently we need to develop techniques that work for broad class of problems. Combinatorial optimization problems provide an example of such a class of problems. Moreover in many emerging data analysis applications, large graphs are defined based on implicit relationship between objects [4, 21]. Subsequently, the goal is to find suitable combinatorial structure in this large implicit graph, e.g., maximum $b$-matchings were considered in [21]. Such edges are often generated through "black box" transducers which have ill understood structure (or are based on domain specific information) and are prohibitive to store explicitly. Therefore in either case, whether the edges are explicitly provided as input or are implicit, it is an useful goal to design algorithms and techniques for graph problems, and in particular combinatorial optimization problems, without storing the edges. The reader would immediately observe the connection to "in place algorithms", which also poses the question of solving a problem using as small a space as possible excluding the input. In many massive data settings, or when the input is implicitly defined, we are faced with the task of designing in place algorithms with no random access or writes to the input. Multipass streaming algorithms seem well placed to answer these types of questions; putting together the two threads of graph streaming and of combinatorial optimization problems. it is natural to ask: *how well can we solve maximum matching problem and its variants using small additional (to the input) space where we only make a few passes over the (possibly adversarial) list of edges?*

*Department of Computer and Information Science, University of Pennsylvania, USA. Email {kookjin,sudipto}@cis.upenn.edu. Research supported in part by an NSF Award CCF-0644119, IIS-0713267 and a gift from Google.

Graph problems were one of the early problems considered in the streaming model, and it was shown that even simple problems such as determining the connectedness of a graph requires $\Omega(n)$ space [19] (throughout this paper $n$ will denote the number of vertices and $m$ will denote the number of edges). This result holds even if a constant number of passes were allowed. However, for problems which are even slightly more involved than connectivity, it is often not clear how to solve them in space $O(C_\varepsilon n \, poly \log n)$, that is, even if we allow the space to be larger than the lower bound by a polylogarithmic factor and allow the constants in the algorithm to depend on an accuracy parameter $0 < \varepsilon \ll 1$. Observe that this space bound is still sublinear in the in the size of the input stream (of $m$ edges).

The semi-streaming model [10, 25] has emerged to be a model of choice in the context of graph processing – by allowing $O(C_\varepsilon n \, poly \log n)$ space for an input stream of $m$ edges defining a graph over $n$ vertices, arriving in any (including adversarial) order. In recent years there have been several new results in this semi-streaming model, for example see [10, 11, 23, 1, 8, 7]. Several of these papers address fundamental graph problems including matchings. These papers demonstrate a rich multidimensional tradeoff between the quality of the solution, the space required and the number of passes over the data (and of course, the running time). Many of these results are likely to be used as a building block for other algorithms. Yet, as with many emerging models, it is natural to ask: *are there broad techniques that can be adapted to the semi-streaming model?*

**Our results:** In this paper we answer both the questions posed previously in the affirmative. In particular we investigate primal–dual based algorithms for solving a subclass of linear programming problems on graphs. The maximum weighted matching (MWM) is a classic example of such. Although augmentation based techniques exist for matching problems in the semi-streaming model, they become significantly difficult in the presence of weights (since we need to find shortest augmenting paths to avoid creating negative cycles) and the best previous result for the maximum weighted matching problem is a $\frac{1}{2} - \varepsilon$ approximation using $O(\frac{1}{\varepsilon^3})$ passes [23] for any $\varepsilon > 0$. Note that the input for weighted problems in the semi-streaming model is a sequence of tuples $\{(i, j, w_{ij})\}$ and the weights do not have to be stored. Since the maximum weighted matching problem is one of the most celebrated combinatorial optimization problems that can be solved optimally, it is natural to ask if we can achieve an efficient approximation scheme, that is, an approximation ratio of $(1-\varepsilon)$ for any $\varepsilon > 0$? The use of linear programming relaxation allows us to design such a approximation scheme, as well as improve the number of passes. See Table 1 for a summary of the results in this paper. We also improve the number of passes for finding the maximum cardinality matching (MCM) in bipartite graphs by a significant amount. The technique extends to several variants such as the $b$-matching problem and matching in general graphs. However, the results for general graphs in this paper have unappealing running times, such as $n^{O(\frac{1}{\varepsilon})}$.

**Subsequent Results:** The question of designing an FPTAS (an approximation scheme where the running time is polynomial in both $n, \frac{1}{\varepsilon}$) for matching problems in general graphs, while using a small number of passes was left open in this paper. In a recent article [2] we show that such a result is possible using a slightly augmented version of the fractional packing framework of [28] (which allows us to control tight sets), based on the ideas of Cunningham-Marsh proof of the laminarity of tight sets [31]. The result in [2] shows that the non-bipartite matching problem reduces (after many non-trivial steps, including finding minimum odd cuts in a space efficient manner) to an LP which corresponds to bipartite matching using "effective weights", and uses the result in this paper for that part. The main ideas and techniques in [2] are orthogonal to this paper.

| Problem | Approx. | No. of Passes $\tau$ | space | paper |
|---|---|---|---|---|
| Bipartite MCM (see all below) | $\frac{2}{3}(1-\varepsilon)$ | $O(\varepsilon^{-1}\log\varepsilon^{-1})$ | | [10] |
| | $1-\varepsilon$ | $O(\varepsilon^{-5})$ | | [7] |
| | $1-\varepsilon$ | $O(\frac{1}{\varepsilon^2}\log\log\frac{1}{\varepsilon})$ | $O(n'(\tau+\log n'))$ | **here** |
| MCM (see MWM below) | $1/2$ | $1$ | | |
| | $1-\varepsilon$ | $\left(\frac{1}{\varepsilon}\right)^{1/\varepsilon}$ | | [23] |
| Bipartite MWM | $1-\varepsilon$ | $O(\varepsilon^{-2}\log\varepsilon^{-1})$ | $O\left(n\left(\tau+\frac{\log n}{\varepsilon}\right)\right)$ | **here** |
| Bipartite $b$-Matching | $1-\varepsilon$ | $O(\varepsilon^{-3}\log n)$ | $\tilde{O}\left(\frac{B}{\varepsilon^3}\right)$ | **here** |
| MWM | $1/6$ | $1$ | $O(n)$ | [10] |
| | $1/5.82$ | $1$ | | [23] |
| | $1/5.58$ | $1$ | | [35] |
| | $1/4.91$ | $1$ | | [9] |
| | $\frac{1}{2}(1-\varepsilon)$ | $O(\varepsilon^{-3})$ | | [23] |
| | $\frac{2}{3}(1-\varepsilon)$ | $O(\varepsilon^{-2}\log\varepsilon^{-1})$ | $O\left(n\left(\tau+\frac{\log n}{\varepsilon}\right)\right)$ | **here** |
| | $1-\varepsilon$ * | $O(\varepsilon^{-4}\log n)$ | $O\left(n\left(\tau+\frac{\log n}{\varepsilon}\right)\right)$ | **here** |
| | $1-\varepsilon$ | $O(\varepsilon^{-7}\log\varepsilon^{-1}\log n)$ | $O(\frac{n\tau}{\varepsilon})$ | [2] |

Table 1: Summary of results: The required time is $O(m\operatorname{poly}(\frac{1}{\varepsilon},\log n))$ for all results, except *. The space bounds of results presented elsewhere were not always obvious, and we have omitted reporting them. Note $n'=\min\{n,|OPT|\log\frac{1}{\varepsilon}\}$ and $B=n$ for the uncapacitated case; otherwise $B=\sum_i b_i$. Please note that the result in [2] is subsequent to this paper and builds on the results herein.

**Our Techniques:** The matching problem has a rich literature, see [6, 16, 20, 24], as well as fast, near linear time approximation algorithms [22, 29, 33, 27, 5]. However, these results use random access significantly and do not translate to results in the semi-streaming model, and newer ideas were used in [10, 23, 35, 9, 8, 7] to achieve results in the semi-streaming model. To improve upon the results in these papers, we need new and more powerful techniques.

In this paper we use the multiplicative-weights update meta-method surveyed in [3]. Over many years there has been a significant thrust in designing fast approximation schemes for packing and covering type linear programming problems [28, 34, 13, 17, 12, 3], to name a few. Such a thrust has existed even outside of theoretical computer science, see the excellent survey in [3]. The meta-method uses the oracle to progressively improve the feasibility of the dual linear program, but uses a (guessed) value of the optimal solution. If the oracle does not fail to provide these improvements within in a predetermined number of iterations, we are guaranteed an approximately feasible dual solution. If the failure of the oracle can be appropriately modified and interpreted to give us a feasible primal solution, we can use that to verify the guess of the optimal solution and as a consequence have an overall scheme. While the key intuition in this paper can be viewed as designing a "streaming separation oracle", it is not clear how to implement (or even define) such an oracle. There are a super-linear (in $n$) number of conditions (constraints, verification of various assumptions) involving the input that need to be satisfied (even though the number of variables are $n$) which mandate random access. Designing an efficient separation oracle is not always trivial even without any constraint on space — one of the interesting contributions of our paper is to show that semi-streaming algorithms for maximal matching can be bootstrapped to achieve a near-optimal matching within a few iterations. However, even if we could design an efficient oracle, the overall scheme to obtain a good semi-streaming algorithm faces a number of roadblocks. First, the multiplicative update method typically requires super-constant number of iterations (to prove feasibility) — this translates to super-constant number of passes. Reducing the number of passes to a constant requires that we recursively identify small and critical subgraphs. Second, for the weighted variants, it is non-trivial to simultaneously ensure that enough global progress is being made per pass, yet the computation in a pass is local (and in small space). Given the fundamental nature of these roadblocks, we expect the different ideas developed herein will find use in other settings as well.

**Other Related Work:** The result in [26], is related but somewhat orthogonal to our discussion in this paper.

**Roadmap:** We revisit the multiplicative weights update method in Section 2. We then demonstrate the simplest possible (but suboptimal in space and the number of passes) application of this framework in Section 3, but in this process we develop the basic oracles. We subsequently show in Section 4 how to (i) improve the space requirement by "simulating" multiple guesses of the optimum solution in parallel as well as (ii) reduce the number of passes by "simulating" multiple iterations of the multiplicative weights update method in a single pass. We show how to remove the dependency on $n$ in Section 5. We finally show some extensions of the maximum matching problem in Section 6 which also demonstrate the generality of our approach.

## 2 The Multiplicative Weights Update Meta-Method

In this section, we briefly explain the multiplicative weights update method; we follow the discussion presented by Arora, Hazan, and Kale [3]. Suppose that we are given the following LP, its dual LP, and a guess of the optimal solution $\alpha$, where $\mathbf{A} \in \mathbb{R}^{n \times m}, \mathbf{b} \in \mathbb{R}^n, \mathbf{c} \in \mathbb{R}^m$:

$$\text{LP:} \begin{cases} \min & \mathbf{b}^\mathrm{T}\mathbf{x} \\ \text{s.t} & \mathbf{A}^\mathrm{T}\mathbf{x} \geq \mathbf{c}, \quad \mathbf{x} \geq \mathbf{0} \end{cases} \qquad \text{Dual LP:} \begin{cases} \max & \mathbf{c}^\mathrm{T}\mathbf{y} \\ \text{s.t} & \mathbf{A}\mathbf{y} \leq \mathbf{b}, \quad \mathbf{y} \geq \mathbf{0} \end{cases}$$

The algorithm proceeds along the weak separation framework [18]. Suppose that the optimal solution is $\alpha$. The **violation** of dual constraint $i$ is $\mathbf{A}_i \mathbf{y} - b_i$. The complementary slackness conditions mandate that for an optimal solution the $x_i(\mathbf{A}_i \mathbf{y} - b_i) = 0$. One way to express the complementary slackness conditions into a single condition is to interpret the primal variables (which are always maintained as positive) as probabilities, and ask: *Is there a vector $\mathbf{y}$ which satisfies $\mathbf{c}^T \mathbf{y} = \alpha$, such that the expected dual violation is at most $\delta$?* The vector $\mathbf{y}$, which is the answer to the question, is termed as a dual witness.

If the answer to the question posed to an oracle is "yes", and the probabilities were chosen such that constraints which had larger violations had larger probability mass; then we have a direction in which the feasibility of the dual solution can be improved. The improvement is measured by a potential function, which is akin to the notion of dual gap.

If the answer is "no" (referred to as the failure of the oracle) — then we know that there is no "good direction" to improve the solution. This serves as a certificate that the dual LP (with the additional constraint that the dual solution is at least $\alpha$) is not feasible. For example, a feasible primal solution which is less than $\alpha$ can be one such certificate. However since we are asking questions to the oracle that have an approximation parameter, the certificate is also approximate at best.

But, note that, neither of the above does not give us a solution to the dual. *However we can produce a dual solution if we achieve two things, in addition to designing an oracle.*

- First, if we are careful in choosing the probabilities (which is what the multiplicative update framework achieves), then we have a way to extract a dual solution which approximately satisfies all the dual constraints. In fact the solution will be the average of the dual witnesses found, and this average will approximately satisfy the dual constraints. It is easy to see that the average satisfies $\mathbf{c}^T\mathbf{y} = \alpha$. Now in many situations, and for the problems in this paper, a simple scaling (multiplying each coordinate by of this average vector by a constant $c$) can ensure dual feasibility and we have a $c$ approximate dual solution.

- Second, note that the approximation also depends on the appropriate guess of $\alpha$. Therefore we need a way of verifying the guess $\alpha$. In this paper we will achieve this by creating a primal feasible solution which is at most $(1 + O(\delta))\alpha$. Observe that the value of a feasible primal solution (minimization) is an upper bound of any feasible dual solution (maximization). Therefore we need to focus on the largest guess of $\alpha$ for which the oracle has not failed.

Since this paper is regarding the application of the multiplicative weights update method in streaming and not about the framework itself, we refer the reader to the original article of [3] for further discussion of the intuition behind the framework. In what follows, we provide a brief review of the main definitions and notation (Definition 1), the meta-algorithm (Algorithm 1), and a restatement of the main result (Theorem 1) in [3]. We also require a minor extension (Corollary 2), and therefore we restate the proof of the main result in [3] for the sake of completeness.

---

**Algorithm 1** The Multiplicative Weights Update Meta-Method [3]

---

1: $u_i^1 = 1$ for all $i \in [n]$.
2: **for** $t = 1$ **to** $T$ **do**
3:     Given $\mathbf{u}^t$, the oracle returns an admissible dual witness $\mathbf{y}^t$. Note that $\mathbf{y}^t$ is not required to be feasible.
4:     Let $\mathrm{M}(i, \mathbf{y}^t) = \mathbf{A}_i \mathbf{y}^t - b_i$ (for all $i$).
5:     For all $i$, set $u_i^{t+1} = \begin{cases} u_i^t(1+\epsilon)^{\mathrm{M}(i,\mathbf{y}^t)/\rho} & \text{if } \mathrm{M}(i, \mathbf{y}^t) \geq 0 \\ u_i^t(1-\epsilon)^{-\mathrm{M}(i,\mathbf{y}^t)/\rho} & \text{if } \mathrm{M}(i, \mathbf{y}^t) < 0 \end{cases}$
6: **end for**
7: Output $\tilde{\mathbf{y}} = \left(\min_i \frac{b_i}{b_i + 4\delta}\right) \frac{1}{T} \sum_t \mathbf{y}^t$. Note, for use in this paper $b_i \geq 1$. This step is dependent on the specific problem.

---

**Definition 1.** *The Algorithm 1 proceeds in iterations and in iteration $t$ finds a dual witness $\mathbf{y}_t$. We define $\mathrm{M}(i, \mathbf{y}^t) = \mathbf{A}_i\mathbf{y}^t - b_i$ to be the **violation** for dual constraint $i$ in iteration $t$. The **expected violation** $\mathrm{M}(\mathcal{D}^t, \mathbf{y}^t)$ is the expected value of $\mathrm{M}(i, \mathbf{y}^t)$ when choosing $i$ with probability proportional to $u_i^t$, i.e., $\sum_i \frac{u_i^t}{\sum_j u_j^t} \mathrm{M}(i, \mathbf{y}^t)$. The dual witness $\mathbf{y}^t$ is defined to be **admissible** if it satisfies*

$$\mathrm{M}(\mathcal{D}^t, \mathbf{y}^t) \leq \delta, \quad \mathbf{c}^\mathrm{T}\mathbf{y}^t \geq \alpha, \quad \text{and} \quad \mathrm{M}(i, \mathbf{y}^t) \in [-\ell, \rho] \quad \forall i \in [n] = \{1, \ldots, n\}$$

*for parameters of the oracle $\ell$ and $\rho$ such that $0 < \ell \leq \rho$ The parameters $\ell, \rho$ will be constants for the oracles in this paper; $\rho$ is called the width parameter of the oracle. The parameters $\epsilon$ and $T$ depend on $\rho$, $\ell$, and $\delta$. Note that admissibility does not imply feasibility.*

**Theorem 1** (A slight rewording of Corollary 3 in [3])**.** *Let $\delta > 0$ be an error parameter and $\epsilon = \min\{\frac{\delta}{4\ell}, \frac{1}{2}\}$. Suppose that the oracle returns an admissible solution (See Definition 1) for $T = \frac{2\rho \ln(n)}{\delta \epsilon}$ iterations in Algorithm 1, then for any constraint $1 \leq i \leq n$ we have: $(1-\epsilon)\sum_t M(i, \mathbf{y}^t) \leq \delta T + \sum_t M(\mathcal{D}^t, \mathbf{y}^t)$. Moreover $\tilde{\mathbf{y}}$ is a feasible solution of the Dual LP.*

*Proof.* We analyze the algorithm using a potential function $\Phi^t = \sum_j u_j^t$. Let $\Upsilon_i^t = u_i^t/\Phi^t$. We assume we have an upper bound $\Psi_i \geq \Upsilon_i^t$. Note that $\Upsilon_i^t$ and $\Psi_i$ are not used in [3]. In this theorem, we use $\Psi_i = 1$ for all $i$ — which is obvious from the fact that all weights $u_i^t$ are positive. We will use a smaller value of $\Psi_i$ to strengthen the theorem later.

We rewrite the proof of [3] using $\Upsilon_i^t$ and $\Psi_i$. Observe that $(1-\epsilon)^{-x}$ and $(1+\epsilon)^x$ are convex (in $x$) for $0 < \epsilon \leq \frac{1}{2}$. Therefore it follows that

$$(1-\epsilon)^{-x} \leq (1+\epsilon x) \quad \text{for } x \in [-1, 0] \qquad \text{and} \qquad (1+\epsilon)^x \leq (1+\epsilon x) \quad \text{for } x \in [0,1]$$

since equality is achieved at the respective endpoints ($x = -1, 0$ for the first fact and $x = 0, 1$ for the second fact). From $M(i, \mathbf{y}^t)/\rho \in [-1, 1]$ (notice $\ell \leq \rho$) and the above facts we have:

$$\begin{aligned}
\Phi^{t+1} &= \sum_i u_i^{t+1} = \sum_{i: M(i, \mathbf{y}^t) < 0} u_i^t (1-\epsilon)^{-M(i, \mathbf{y}^t)/\rho} + \sum_{i: M(i, \mathbf{y}^t) \geq 0} u_i^t (1+\epsilon)^{M(i, \mathbf{y}^t)/\rho} \\
&\leq \sum_i u_i^t (1 + \epsilon M(i, \mathbf{y}^t)/\rho) = \Phi^t + \frac{\epsilon}{\rho} \sum_i u_i^t M(i, \mathbf{y}^t) = \Phi^t + \frac{\epsilon \Phi^t}{\rho} \sum_i \frac{u_i^t}{\Phi^t} M(i, \mathbf{y}^t) \\
&= \Phi^t (1 + \epsilon M(\mathcal{D}^t, \mathbf{y}^t)/\rho) \qquad \left(\text{Using the definition of } M(\mathcal{D}^t, \mathbf{y}^t)\right) \leq \Phi^t e^{\left(\epsilon M(\mathcal{D}^t, \mathbf{y}^t)/\rho\right)}.
\end{aligned}$$

Therefore we can conclude that, $\Phi^{T+1} \leq \Phi^1 e^{\left(\epsilon \sum_{t=1}^T M(\mathcal{D}^t, \mathbf{y}^t)/\rho\right)}$. From the algorithm and the definitions of $\Psi_i$,

$$u_i^1 (1+\epsilon)^{\left(\sum_{t: M(i, \mathbf{y}^t) \geq 0} M(i, \mathbf{y}^t)/\rho\right)} \cdot (1-\epsilon)^{-\left(\sum_{t: M(i, \mathbf{y}^t) < 0} M(i, \mathbf{y}^t)/\rho\right)} = u_i^{T+1} \leq \Phi^{T+1} \Psi_i. \qquad (1)$$

From the definition of $\Phi^t, \Upsilon_i^t, \Psi_i$, we get $\Phi^1 = u_i^1/\Upsilon_i^1$. Using $\Phi^{T+1}/\Phi^1$ in equation (1), we get,

$$u_i^1 (1+\epsilon)^{\left(\sum_{t: M(i, \mathbf{y}^t) \geq 0} M(i, \mathbf{y}^t)/\rho\right)} \cdot (1-\epsilon)^{-\left(\sum_{t: M(i, \mathbf{y}^t) < 0} M(i, \mathbf{y}^t)/\rho\right)} \leq \frac{\Psi_i}{\Upsilon_i^1} u_i^1 e^{\left(\epsilon \sum_{t=1}^T M(\mathcal{D}^t, \mathbf{y}^t)/\rho\right)}.$$

Applying the natural log function and simplifying we get:

$$\ln(1+\epsilon) \sum_{t: M(i, \mathbf{y}^t) \geq 0} M(i, \mathbf{y}^t) - \ln(1-\epsilon) \sum_{t: M(i, \mathbf{y}^t) < 0} M(i, \mathbf{y}^t) \leq \rho \ln \frac{\Psi_i}{\Upsilon_i^1} + \epsilon \sum_{t=1}^T M(\mathcal{D}^t, \mathbf{y}^t).$$

Now $\ln(1+\epsilon) - \epsilon(1-\epsilon) \geq 0$; we have equality at $\epsilon = 0$ and the first derivative of the left hand side with respect to $\epsilon$ is positive for $\epsilon > 0$. Likewise (using the derivative, but only over the range $0 < \epsilon \leq \frac{1}{2}$) we have $\ln(1-\epsilon) + \epsilon(1+\epsilon) \geq 0$. Therefore using $\ln(1+\epsilon) \geq \epsilon(1-\epsilon)$ and $\ln(1-\epsilon) \geq -\epsilon(1+\epsilon)$ we get:

$$\frac{\rho}{\epsilon} \ln \frac{\Psi_i}{\Upsilon_i^1} + \sum_{t=1}^{T} M(\mathcal{D}^t, \mathbf{y}^t) \geq (1-\epsilon) \sum_{t:M(i,\mathbf{y}^t)\geq 0} M(i, \mathbf{y}^t) + (1+\epsilon) \sum_{t:M(i,\mathbf{y}^t)<0} M(i, \mathbf{y}^t)$$

$$= (1-\epsilon) \sum_{t=1}^{T} M(i, \mathbf{y}^t) + 2\epsilon \sum_{t:M(i,\mathbf{y}^t)<0} M(i, \mathbf{y}^t)$$

$$\geq (1-\epsilon) \sum_{t=1}^{T} M(i, \mathbf{y}^t) - 2\epsilon \ell T \qquad \left(\text{From } M(i, \mathbf{y}^t) \geq -\ell\right)$$

Selecting $T = \frac{2\rho}{\delta \epsilon} \ln \frac{\Psi_i}{\Upsilon_i^1}$, $\Psi_i = 1$, and $\epsilon = \min\left\{\frac{\delta}{4\ell}, \frac{1}{2}\right\}$, we obtain $(1-\epsilon) \sum_t M(i, \mathbf{y}^t) \leq \delta T + \sum_t M(\mathcal{D}^t, \mathbf{y}^t)$. Note that $\ln(n)$ arises from $\ln \frac{\Psi_i}{\Upsilon_i^1}$. Since $\frac{1}{T} \sum_t M(i, \mathbf{y}^t) = M(i, \frac{1}{T} \sum_t \mathbf{y}^t)$ and $M(\mathcal{D}^t, \mathbf{y}^t) \leq \delta$ for all $t$, we have $M(i, \frac{1}{T} \sum_t \mathbf{y}^t) \leq (1-\epsilon)^{-1}(2\delta) \leq 4\delta$ (dividing both the left and right side of the inequality by $T$) or $\mathbf{A}_i \left(\frac{1}{T} \sum_t \mathbf{y}^t\right) \leq b_i + 4\delta$. This translates to the fact that $\tilde{\mathbf{y}}$ satisfies $\mathbf{A}_i \tilde{\mathbf{y}} \leq b_i$. □

We also obtain Corollary 2, since we will assign different initial weights $u_i^1$ on constraints and use different values of $\Upsilon_i$. In Section 5, we eliminate the dependency on $\ln(n)$ in Theorem 1 using the Corollary.

**Corollary 2.** *Let $\delta > 0$ be an error parameter and $\epsilon = \min\{\frac{\delta}{4\ell}, \frac{1}{2}\}$. Let $\Upsilon_i^t = u_i^t / \sum_j u_j^t$ and let $\Psi_i \geq \Upsilon_i^t$ for all $t$. Suppose that the oracle returns an admissible solution (See Definition 1) for $T = \frac{2\rho}{\delta \epsilon} \ln \max_i \frac{\Psi_i}{\Upsilon_i^1}$ iterations in Algorithm 1, then for any constraint $1 \leq i \leq n$ we have: $(1-\epsilon) \sum_t M(i, \mathbf{y}^t) \leq \delta T + \sum_t M(\mathcal{D}^t, \mathbf{y}^t)$. Moreover $\tilde{\mathbf{y}}$ is a feasible solution of the Dual LP.*

# 3 Warming Up: $O(\frac{1}{\varepsilon^3} \log n)$-pass Algorithms

In this section, we provide a $(1-\varepsilon)$-approximation algorithm for bipartite MCM and MWM that uses $O(\frac{1}{\varepsilon^3} \log n)$ passes. We will use the multiplicative weights update method reviewed in Section 2. Recall that the method provides a solution the dual problem. We formulate the primal LP (LP1 and LP3) to be the dual of the actual LP for MCM and MWM (LP2 and LP4) respectively. Note that the edges are undirected in these LP formulations.

$$\begin{array}{ll} \min & \sum_i x_i \\ \text{s.t} & x_i + x_j \geq 1 \quad \forall (i,j) \in E \\ & x_i \geq 0 \quad \forall i \in V \end{array} \quad \text{(LP1)} \qquad \begin{array}{ll} \max & \sum_{(i,j) \in E} y_{ij} \\ \text{s.t} & \sum_{j:(i,j) \in E} y_{ij} \leq 1 \quad \forall i \in V \\ & y_{ij} \geq 0 \quad \forall (i,j) \in E \end{array} \quad \text{(LP2)}$$

$$\begin{array}{ll} \min & \sum_i x_i \\ \text{s.t} & x_i + x_j \geq w_{ij} \quad \forall (i,j) \in E \\ & x_i \geq 0 \quad \forall i \in V \end{array} \quad \text{(LP3)} \qquad \begin{array}{ll} \max & \sum_{(i,j) \in E} w_{ij} y_{ij} \\ \text{s.t} & \sum_{j:(i,j) \in E} y_{ij} \leq 1 \quad \forall i \in V \\ & y_{ij} \geq 0 \quad \forall (i,j) \in E \end{array} \quad \text{(LP4)}$$

The integrality gap of LP2 (and LP4) is one, since we have a bipartite graph [31]. We first present an algorithm for MCM and then generalize the algorithm for MWM.

## 3.1 The Simple Case of MCM

We apply the multiplicative weights update method [3] with the oracle provided in Algorithm 2. Recall that if the oracle does not fail, Algorithm 1 returns a feasible solution for LP2 after $T$ iterations.

We also make the observation that: we can compute a maximal matching in one pass in the semi-streaming model in $O(m)$ time and $O(n)$ space. It is trivial to observe that any maximal matching is a 2 approximation to the maximum cardinality matching.

---
**Algorithm 2** Oracle for LP1. The input is $\{u_i^t\}_{i \in V}, \alpha$.

---
1: Let $x_i = \frac{\alpha}{\sum_j u_j^t} u_i^t$. Let $E_{violated} = \{(i,j) | x_i + x_j < 1\}$.
2: Find a **maximal matching** $S$ in $E_{violated}$. Let $\Delta = |S|$.
3: **if** $\Delta < \delta\alpha$ **then**
4:     For each $(i,j) \in S$, increase $x_i$ and $x_j$ by 1. Observe that **x** is feasible for LP1.
5:     Return **x** and report failure.
6: **else**
7:     Return $y_{ij} = \alpha/\Delta$ for $(i,j) \in S$ and $y_{ij} = 0$ otherwise.
8: **end if**

---

**Lemma 3.** *If $\Delta \geq \delta\alpha$, the oracle described in Algorithm 2 returns an admissible solution with $\ell = 1$ and $\rho = 1/\delta$.*

*Proof.* Note $y_{ij} = \alpha/|S|$ for $(i,j) \in S$ and $y_{ij} = 0$ for $(i,j) \notin S$. Therefore, it is obvious that $\sum_{(i,j) \in E} y_{ij} = \alpha$. Since the vector **c** is all 1, we have $\mathbf{c}^T \mathbf{y} \geq \alpha$.

For each edge $(i,j) \in S$ we have $x_i + x_j < 1$, and therefore we have $\sum_{(i,j) \in S} y_{ij}(x_i + x_j) < \alpha$. Therefore $\sum_{(i,j) \in E} y_{ij}(x_i + x_j) = \sum_{(i,j) \in S} y_{ij}(x_i + x_j) < \alpha$. This rewrites to $\sum_i x_i \sum_{j:(i,j) \in E} y_{ij} < \alpha$. Observe that $\sum_i x_i = \alpha$ and therefore $\sum_i x_i (\sum_{j:(i,j) \in E} y_{ij} - 1) < 0$.

Thus $M(\mathcal{D}^t, \mathbf{y}) = \frac{1}{\alpha} \sum_i x_i M(i, \mathbf{y}) \leq 0 < \delta$ and the solution is admissible. Now $M(i, \mathbf{y}) = \sum_{j:(i,j) \in E} y_{ij} - 1 \geq -1$. Since $S$ is a matching, for every $i$ at most one $y_{ij} \neq 0$ and moreover $y_{ij} \leq 1/\delta$ (otherwise the oracle has failed). Therefore $-1 \leq M(i, \mathbf{y}) \leq 1/\delta$. □

**Lemma 4.** *If $\Delta < \delta\alpha$, Algorithm 2 returns a feasible solution for LP1 with value at most $(1+2\delta)\alpha$.*

*Proof.* Consider $(i,j) \in E$ such that $x_i + x_j < 1$. Since $S$ was maximal, there exists an edge in $S$ that is adjacent to either $i$ or $j$. So $x_i$ or $x_j$ is increased by at least 1 and the constraint corresponding to edge $(i,j)$ is satisfied. For each edge $(i', j') \in S$, we increase the objective value by 2 and $|S| \leq \delta\alpha$. Since we started with $\sum_i x_i = \alpha$, the solution returned has value at most $(1 + 2\delta)\alpha$ after the increase. □

**Theorem 5.** *For any $\varepsilon \leq \frac{1}{2}$ let $T = O(\frac{1}{\varepsilon^3} \log n)$. Using $T + 1$ passes and space $O(\frac{nT}{\varepsilon})$ and time $O(\frac{mT}{\varepsilon})$ time we can find a $(1 - \varepsilon)$ approximation to the maximum cardinality matching in bipartite graphs. This implies a $\frac{2}{3}(1 - \varepsilon)$ result for general graphs using the integrality gap results of [14, 15].*

*Proof.* We use the first pass to compute $OPT$ to within factor 2 — this follows from the fact that any maximal matching is a 2 approximation to the maximum cardinality matching. Suppose the size of the maximal matching we found is $q$. We try all possible values of $\alpha = (1 + \frac{\varepsilon}{3})^j q$ where $j \geq 0$ and $\alpha \leq 2q(1 + \frac{\varepsilon}{3})$ in parallel. This corresponds to $O(\frac{1}{\varepsilon})$ guesses of $\alpha$.

Let $\delta = \varepsilon/12$. Therefore $\epsilon = \min\{\frac{\delta}{4}, \frac{1}{2}\} = \varepsilon/48$ since $\ell = 1$. Note, the parameters $\epsilon, \varepsilon$ are different. We now apply the Algorithm 1 using Algorithm 2 as the oracle.

Let $\alpha_0$ to be the smallest value of $\alpha$ which is above $OPT$, i.e., $\alpha_0 \geq OPT > \alpha_0/(1+\frac{\varepsilon}{3})$. Consider $\alpha \leq \alpha_0/(1+\frac{\varepsilon}{3})^2 < OPT/(1+\frac{\varepsilon}{3})$. For any such value of $\alpha$ it is impossible that the oracle fails since we return a feasible primal solution of value at most $(1+2\delta)\alpha = (1+\varepsilon/6)\alpha \leq (1+\frac{\varepsilon}{6})OPT/(1+\frac{\varepsilon}{3}) < OPT$. Therefore if we consider the largest value of $\alpha$ for which we **do not** return a feasible primal solution, that value must satisfy $\alpha \geq \alpha_0/(1 + \frac{\varepsilon}{3})^2$. Let this value be $\alpha^*$. Using Theorem 1, after $T$ iterations we have a feasible dual solution $\tilde{\mathbf{y}}$. Note all $b_i = 1$ and $\delta = \varepsilon/3$. By construction $\sum_{(i,j) \in E} y_{ij}^t = \alpha^*$ for every $t$. Therefore

$$\sum_{(i,j) \in E} \tilde{y}_{ij} \geq \frac{\alpha^*}{1 + \frac{\varepsilon}{3}} \geq \frac{\alpha_0}{(1 + \frac{\varepsilon}{3})^3} \geq (1-\varepsilon)\alpha_0 \geq (1-\varepsilon)OPT$$

The time and space bounds follow easily. To find the actual matching: Let $\varepsilon = \varepsilon'/2$ and we run the above steps to find a fractional solution. We find $T$ matchings before we return the best fractional solution, there are at most $m' = O(nT)$ non-zero entries in the solution. Focus on the graph $G'$ defined by these edges only. The fractional solution of the original graph remains a fractional solution in $G'$. We now have random access to these edges in $G'$ and can find a $(1-\varepsilon)$ approximation to the best matching contained in these edges (which is at least the same value as the fractional solution, we use the integrality of the bipartite matching polytope) in time $\tilde{O}(m')$ using known algorithms [20, 24, 22]. The overall approximation is $(1-\varepsilon)^2 \geq (1-\varepsilon')$. □

## 3.2 Abstracting the Oracle

The intuition behind the oracle, Algorithm 2 will be used for all the algorithms. Although it is not difficult to see that the discussion about the oracle need not be limited to linear programs for matching, we do not diverge from that topic in the interest of brevity. In Algorithm 2 we must choose a subset $S$ of edges which balances two critical properties:

**Admissibility** : Each vertex $i$ is adjacent to at most one edge in $S$. The weights assigned to the edges in $S$ (note, they are identical for a specific iteration $t$) define the parameters $\ell, \rho$. These parameters determine the number of iterations.

**Verification** : Focusing on the violations in the primal solution allows us to produce a feasible primal solution and verify $\alpha$. For each violated edge $(i,j)$ in the primal solution, we pick at least one adjacent edge.

Any **maximal** matching in $E_{violated}$ satisfies both conditions. Since we consider the violated edges only, the algorithm is natural. Observe that the multiplicative framework operates on dual violations whereas the oracle operated on primal violations. In a sense, the problem of finding the maximum matching problem in bipartite graphs reduces to the problem of repeatedly finding maximal matchings in subgraphs defined by primal violations (corresponding to the edges). These violations can be easily defined by a simple filtering conditions, for example, does the input edge satisfy $x_i + x_j < 1$ using the current solution $\mathbf{x}$ of the primal, and can be implemented in the semi-streaming model. We now proceed to discuss weighted graphs — observe that weights will also arise naturally in the unweighted case as we improve Theorem 5.

## 3.3 The Not So Simple Case of MWM

Note that the input for weighted problems in the semi-streaming model is a sequence of tuples $\{(i,j,w_{ij})\}$ and the weights do not have to be stored. We can easily compute a maximum matching in a single pass using $O(n)$ space and $O(m)$ time. It is shown in [10] that we can compute a $1/6$ approximation to the maximum weighted matching in a single pass using $O(n)$ space and $O(m)$ time.

In a weighted graph, the verification condition must be strengthened to handle the complications introduced by edge weights. In LP2, if we increase $x_i$ by 1, then all the edges adjacent to $i$ are satisfied. Therefore if only a few primal constraints were violated then we could produce a primal feasible solution which is close to $\alpha$. It is not true in LP4, we now have to increase $x_i$ by the amount of violation. However trying to fix the verification condition by itself does not help, any change also has to ensure the admissibility condition and a larger increase in $x_i$ corresponds to larger $\rho$.

Let $w(S) = \sum_{(i,j) \in S} w_{ij}$ denote the total weight for any set of edges $S$. The oracle will search for a set $S$ of edges which satisfy the following:

**Weighted Admissibility of** $S$ : There exists a matching $S'$ contained in $S$ such that $w(S') = \Omega(w(S))$. We use $S'$ to construct a dual witness. Since $S'$ is a matching we will have some control over $\ell, \rho$.

**Weighted Verification of** $S$ : For each violated edge $(i, j)$, we pick at least one edge adjacent to it whose weight is $\Omega(w_{ij})$. We need all of $S$ to produce an upper bound of the primal solution.

---

**Algorithm 3** Oracle for LP3.

1: Let $x_i = \frac{\alpha}{\sum_j u_j^t} u_i^t$.
2: Let $E_{violated,k} = \{(i,j)|x_i + x_j < w_{ij}, \alpha/2^k < w_{ij} \leq \alpha/2^{k-1}\}$.
3: Find a maximal matching $S_k$ in $E_{violated,k}$ for each $k = 1, \cdots, \lceil \log \frac{n}{\delta} \rceil = O(\log n)$.
4: Let $S = \cup_k S_k, \Delta = w(S)$.
5: **if** $\Delta < \delta \alpha$ **then**
6:     For each $(i,j) \in S$, increase $x_i$ and $x_j$ by $2w_{ij}$.
7:     Further increase every $x_i$ by $\frac{\delta \alpha}{n}$. Return **x** and report failure.
8: **else**
9:     $S' \leftarrow \emptyset$.
10:     **repeat**
11:       Pick a heaviest edge $(i,j)$ from $S$ and add it to $S'$
12:       Eliminate all edges adjacent to $i$ or $j$ from $S$.
13:     **until** $S = \emptyset$
14:     Return $y_{ij} = \alpha/w(S')$ for $(i,j) \in S'$ and $y_{ij} = 0$ otherwise.
15: **end if**

---

In order to satisfy the modified conditions, we partition the edges depending on their weights.

**Definition 2.** *An edge $(i,j)$ is in* **tier** *$k$ if $\alpha/2^k < w_{ij} \leq \alpha/2^{k-1}$.*

Algorithm 3 is the oracle for LP3. Before we prove the admissibility and verification conditions we prove an useful lemma which is a property of the constraints.

**Lemma 6.** *If* **y** *satisfies $\sum_{(i,j) \in E} w_{ij} y_{ij} = \alpha$ then for any weights* $\mathbf{u}^t$*, if we have $x_i^t = \alpha u_i^t/(\sum_j u_j^t)$ then* $\mathrm{M}(\mathcal{D}^t, \mathbf{y}) = \frac{1}{\alpha} \left( \sum_{(i,j) \in E} y_{ij}(x_i^t + x_j^t - w_{ij}) \right)$.

*Proof.* From the definition of $\mathrm{M}(\mathcal{D}^t, \mathbf{y})$ we have:

$$\alpha \mathrm{M}(\mathcal{D}^t, \mathbf{y}) = \sum_i x_i^t ( \sum_{j:(i,j) \in E} y_{ij} - 1) = \sum_{(i,j) \in E} y_{ij}(x_i^t + x_j^t) - \sum_i x_i^t$$

$$= \sum_{(i,j) \in E} y_{ij}(x_i^t + x_j^t) - \sum_{(i,j) \in E} w_{ij} y_{ij} = \sum_{(i,j) \in E} y_{ij}(x_i^t + x_j^t - w_{ij})$$

The lemma follows. □

**Lemma 7.** *(Weighted Admissibility.)* *The matching $S' \subseteq S$ constructed by Algorithm 3 satisfies $w(S') \geq w(S)/5$. As a consequence, if $\Delta \geq \delta\alpha$ the Algorithm 3 returns an admissible dual witness with $\rho = \frac{5}{\delta}$ and $\ell = 1$.*

*Proof.* Observe that we choose at most one edge incident to $i$ from each tier of weight. Consider the matching $S'$ constructed by Algorithm 3. Suppose that $(i,j) \in S'$ is in tier $k$ and consider the edges $A_i = \{(i,j')|(i,j') \in S, j \neq j'\}$ which are eliminated from $S$ by the inclusion of this edge $(i,j)$ in $S'$. Each element in $A_i$ has a lower weight than $(i,j)$; otherwise we would have chosen that eliminated edge instead of $(i,j)$ (the weights cannot be equal since there cannot be any other edge from tier $k$ which is incident on $i$). Therefore the edges in $A_i$ belong to tiers numbered $k+1$ or larger (since they have a lower weight). The weight of any ignored edge in tier $k+q$ can be upper bounded by $w_{ij}/2^{q-1}$. These weights add up to $2w_{ij}$ since we have at most one edge from each tier. Therefore the sum of the weights of the edges in $A_i, A_j$ amount to at most $4w_{ij}$. Summing over all $(i,j) \in S'$, $w(S - S') \leq 4w(S')$ and thus $w(S') \geq w(S)/5$.

For the second part of the lemma, observe that $y_{ij} = \alpha/w(S') \leq 5\alpha/w(S) = 5\alpha/\Delta \leq \frac{5}{\delta}$. The parameter $\ell$ remains 1 due to the same reason as in the proof of Lemma 3. We observe that (since $y_{ij} = 0$ for $(i,j) \notin S'$);

$$\mathbf{c}^T \mathbf{y} = \sum_{(i,j)} w_{ij} y_{ij} = \sum_{(i,j) \in S'} w_{ij} y_{ij} = \sum_{(i,j) \in S'} w_{ij} \frac{\alpha}{w(S')} = \frac{\alpha}{w(S')} w(S') = \alpha$$

Applying Lemma 6 we immediately get $\alpha M(\mathcal{D}^t, \mathbf{y}) = \sum_{(i,j) \in E} y_{ij}(x_i^t + x_j^t - w_{ij})$. Now if $x_i^t + x_j^t - w_{ij} \geq 0$ then $y_{ij} = 0$ by construction. or in other words, $M(\mathcal{D}^t, \mathbf{y}) \leq 0 \leq \delta$. The lemma follows. $\square$

**Lemma 8.** *(Weighted Verification.)* *For every violated edge $(i,j)$ in one of the $O(\log n)$ tiers, we pick at least one edge in $S$ adjacent to $(i,j)$ whose weight is at least $w_{ij}/2$. As a consequence, if $\Delta < \delta\alpha$ then the algorithm returns a feasible primal solution for LP3 with value at most $(1 + 5\delta)\alpha$.*

*Proof.* Suppose that $(i,j)$ is violated in tier $k$. Then, since $S_k$ is a maximal matching, we must have chosen at least one edge in $S_k$ which is adjacent to $i$ or $j$. The weight of that chosen edge in $S_k$ has to be at least $w_{ij}/2$ since the weights of two edges that belong to the same tier differ at most by a factor of 2.

For the second part of the proof we follow the argument in Lemma 4, with one change. Suppose that $x_i + x_j < w_{ij}$ for an edge $(i,j) \in E$. We have two cases, either the edge was chosen in one of the tiers (say $k$) or $w_{ij} \leq \delta\alpha/n$. The second case is easier, since we increase each $x_i$ by at least $\delta\alpha/n$, we definitely satisfy the constraint for $(i,j)$ in this case.

For the first case, observe that in the first part of the lemma we proved that we selected an edge $e \in S$ incident on $i$ or $j$ with weight $w_e \geq w_{ij}/2$. Therefore we increased $x_i$ or $x_j$ by at least $2w_e \geq w_{ij}$. We satisfy the constraint for $(i,j)$ in this case as well. Therefore $\mathbf{x}$ is feasible.

For each edge $(i,j) \in S$, we increase $\sum_i x_i$ by $4w_{ij}$. Therefore over all the edges we increase $\sum_i x_i$ by $4w(S) = 4\Delta$. Since we started with $\sum_i x_i = \alpha$ we have $\sum_i x_i = \alpha + 4\Delta < (1 + 4\delta)\alpha$ after we increase $x_i$ based on the edges. We now have an additional increase in $x_i$ which adds $\delta\alpha$ to $\sum_i x_i$. The lemma follows. $\square$

The rest of the argument is almost identical to MCM and proof of Theorem 5 with four changes: (i) $\rho$ increases to $\frac{5}{\delta}$ from $\frac{1}{\delta}$ (ii) we need to set $\delta = \varepsilon/30$ since the primal feasible solution returned is at most $(1 + 5\delta)\alpha$ (iii) we start with the 1/6 approximation provided by [10] which uses $O(n)$ space and (iv) for the final rounding scheme we use the recent result of [5]. The space bound increases since the oracle now uses $O(n \log n)$ space (and as before we have $O(\frac{1}{\varepsilon})$ oracles being run in parallel).

**Theorem 9.** *For any $\varepsilon \leq \frac{1}{2}$ in $T = O(\frac{1}{\varepsilon^3} \log n)$ passes, and $O(\frac{nT}{\varepsilon} + \frac{n}{\varepsilon} \log n)$ space we can compute a $(1 - \varepsilon)$ approximation for maximum weighted matching in bipartite graphs.*

## 4 Reducing the Space Requirement and the Number of Passes

So far we have not used the fact that we are trying to solve the same LP for different guesses of the parameter $\alpha$. Moreover we have used one pass for each invocation of the oracle. The number of passes is equal to the number of iterations plus one; the first pass is used to guess the values of $\alpha$. In this section, we first reduce the space required to manage the multiple guesses of $\alpha$. Subsequently, we reduce the number of passes by executing multiple iterations of the algorithm in one pass – this can be viewed as making a "step" which is significantly larger than what is provided by the basic analysis in the previous section. We focus on the weighted case.

### 4.1 Reducing the Space Requirement

In what follows we show how to preserve the admissibility condition across different values of the guessed parameter $\alpha$, and run the $O(\frac{1}{\varepsilon})$ guesses (in Theorem 9) in parallel without increasing the space requirement by a factor $O(\frac{1}{\varepsilon})$. The key intuition is that we are trying to find feasible solutions for the same instance of LP4 but different values of the objective function. If in a single iteration we make progress for a large value of $\alpha$ then we also make progress for a smaller value of $\alpha$.

Observe that the proofs of Theorem 5 and 9 use the largest value of $\alpha$ for which we have not produced a feasible primal solution. Suppose that we can prove that we would make the same choices for different values of $\alpha$. Then, when we produce a feasible primal solution for some guess of $\alpha$ (the oracle fails), it may be that for a smaller guess of $\alpha$ the oracle does not fail. We can continue with the smaller guess of $\alpha$, as if the larger guess was never made! Therefore we will avoid running separate oracles for the different guesses of $\alpha$ and thereby save space. We begin with the following definition:

**Definition 3.** *A sequence $\mathbf{y}^1, \mathbf{y}^2, \cdots, \mathbf{y}^t$ is admissible if all $\mathbf{y}$ are admissible when we apply $\mathbf{y}^1, \mathbf{y}^2, \cdots, \mathbf{y}^t$ in the given order.*

**Lemma 10.** *Let $\alpha, \alpha'$ be guesses of the optimal solution with $\alpha > \alpha'$. If a sequence $\mathbf{y}^1, \mathbf{y}^2, \cdots, \mathbf{y}^t$ is admissible for $\alpha$, the sequence is also admissible for $\alpha'$.*

*Proof.* Consider running the two copies of the Algorithm 1: for the values of $\alpha$ and $\alpha'$. Observe that $\mathrm{M}(i, \mathbf{y})$ only depends on $\mathbf{y}$ and therefore the parameters $\ell, \rho$ do not depend on $\alpha, \alpha'$. Moreover the actual weights of the edges do not change and therefore for any vector $\mathbf{y}$ if $\mathbf{c}^T \mathbf{y} \geq \alpha$ then $\mathbf{c}^T \mathbf{y} > \alpha'$.

Therefore to show admissibility, it suffices to prove that $\mathrm{M}(\mathcal{D}^q, \mathbf{y}^q) \leq \delta$ for all $q \leq t$ for the smaller value $\alpha'$ assuming that $\mathbf{y}^q$ satisfied $\mathrm{M}(\mathcal{D}^q, \mathbf{y}^q) \leq \delta$ for all $q \leq t$ for the larger value $\alpha$. We prove this using induction.

Initially $\mathbf{u}$ is same for both copies of the algorithm (as described so far, we have used $u_1^i = 1$, but we will be changing this in the next section). Now $\mathrm{M}(\mathcal{D}^1, \mathbf{y}^1) = \frac{1}{\sum_j u_j^1} \sum_i u_i^1 \mathrm{M}(i, \mathbf{y}^1)$ and is independent of $\alpha$. Therefore $\mathbf{y}^1$ is admissible for $\alpha'$. This proves the base case.

Suppose that we have proven the hypothesis up to $q = k$ and we apply $\mathbf{y}^1, \cdots, \mathbf{y}^k$ to both the algorithms corresponding to $\alpha$ and $\alpha'$. Observe that $\rho, \mathrm{M}(i, \mathbf{y})$ are unchanged and therefore the weights $u_i^{k+1}$ is the same for both $\alpha$ and $\alpha'$. But $\mathrm{M}(\mathcal{D}^{k+1}, \mathbf{y}^{k+1}) = \frac{1}{\sum_j u_j^{k+1}} \sum_i u_i^{k+1} \mathrm{M}(i, \mathbf{y}^{k+1})$ and for all $i$ both algorithms have the same value of $\frac{1}{\sum_j u_j^{k+1}} u_i^{k+1}$ and $M(i, \mathbf{y}^{k+1})$ since these quantities are independent of $\alpha$. Therefore $\mathbf{y}^{k+1}$ is also admissible for $\alpha'$. The lemma follows by induction. □

**The algorithm:** We start with $\alpha$ being the upper bound of the maximum matching. Each time the oracle fails, we reduce $\alpha$ by $(1 + \frac{\varepsilon}{3})$ factor while keeping the weights of constraints and $\{\mathbf{y}^t\}$ fixed. This is possible since the sequence of $\mathbf{y}$ remains admissible with the same width parameter. The total number of successful iterations remains the same but we need an additional iteration for each time the oracle reports failure. However we only have to provision for solving one copy of the oracle.

---

**Algorithm 4** Improved Algorithm for MWM (reducing space).

1: In one pass, find a 6 approximate maximum matching using [10] and let $\alpha_0$ be the weight of the matching.
2: $u_i^1 = 1$ for all $i \in [n]$ and $\alpha = 6\alpha_0$
3: **for** $t = 1$ **to** $T$ **do**
4:   Given $u_i^t$, run the oracle (Algorithm 3).
5:   If the oracle failed decrease $\alpha$ by factor $(1 + \frac{\varepsilon}{3})$ and repeat line 4.
6:   Let $M(i, \mathbf{y}^t) = \mathbf{A}_i \mathbf{y}^t - b_i$. ($\mathbf{y}$ is an admissible dual witness now)
7:   $u_i^{t+1} = \begin{cases} u_i^t(1+\epsilon)^{M(i,\mathbf{y}^t)/\rho} & \text{if } M(i, \mathbf{y}^t) \geq 0 \\ u_i^t(1-\epsilon)^{-M(i,\mathbf{y}^t)/\rho} & \text{if } M(i, \mathbf{y}^t) < 0 \end{cases}$
8: **end for**
9: Output $\frac{1}{T} \frac{1}{1+4\delta} \sum_t \mathbf{y}^t$.

---

We can now show that Theorem 9 holds with space $O(n(T + \log n))$, but uses $T' = T + O(\frac{1}{\varepsilon})$ passes. Formally,

**Theorem 11.** *For any $\varepsilon \leq \frac{1}{2}$ in $T = O(\frac{1}{\varepsilon^3} \log n)$ passes, and $O(n(T + \log n))$ space we can compute a $(1 - \varepsilon)$ approximation for maximum weighted matching in bipartite graphs.*

## 4.2 Reducing the number of passes

Consider the two conditions for the oracle given in the previous section, and for the sake of example, consider the cardinality case. Suppose that we just performed an update based on a dual witness $\mathbf{y}$. Observe that $\mathbf{x}^t = \mathbf{x}(\mathbf{u}^t)$ and for the next step, the admissibility condition ($M(\mathcal{D}^t, \mathbf{y}) \leq 0 \leq \delta$) remains satisfied as long as the edges $(i, j)$ in $\mathbf{y}$ satisfy $x_i^t + x_j^t < 1$. Therefore as a new approach, we do not invoke the oracle again as long as we have such a solution. In other words, we can use the same matching returned as a dual witness for multiple iterations or until one of its edges satisfies the corresponding primal constraint $x_i + x_j \geq 1$.

Therefore it appears that we can simulate multiple iterations in a single pass. But if $x_i + x_j$ is close to 1 then this idea need not be useful because we may satisfy that edge in a single step. Observe that this idea automatically brings up the notion of weights even in the context of MCM. The high-level idea for the oracle is similar to the construction in Section 3 – but there are significant differences and two major issues arise.

- First, we cannot use uniform values for the entries of $\mathbf{y}$ as in Section 3, even in the setting of MCM. Suppose that $S$ contains $(i, j)$ and $(i', j')$ where $1 - x_{i'} - x_{j'}$ is greater than $1 - x_i - x_j$. If we assign large values to $y_{ij}$ and $y_{i'j'}$, it decreases the number of iterations per pass (due to normalization the $x_i$ for the matched edges rise quickly, and we satisfy the constraint). If we assign small values to $y_{ij}$ and $y_{i'j'}$, it increases the total number of iterations and it may also result in inadmissible $\mathbf{y}$, i.e., $\mathbf{c}^T\mathbf{y} < \alpha$.

- Second, we have to modify the verification condition in Section 3 so that the condition handles the values of $w_{ij} - x_i - x_j$ and keep the increase of the solution minimal. For example, (again using the cardinality case as an example) increasing the value of $x_i$ and $x_j$ less than 1 in the verification step can result in an infeasible primal solution. On the other hand, increasing $x_i$ and $x_j$ by 1 can result in a larger approximation factor.

In what follows, we avoid both the issues by defining the tier of an edge based on the violation instead of the edge weight. Moreover we ensure that for different edges $(i,j)$ the $y_{ij}$ values are different — this can be viewed as setting $w_{ij}y_{ij}$ proportional to the violation in $(i,j)$. Therefore the accounting for the admissibility and verification conditions are different.

**Definition 4.** *Define $v_{ij} = w_{ij} - x_i - x_j$ to be the **(primal) violation** of an edge $(i,j) \in E$ (the edge is not violated if $v_{ij} < 0$). An edge $(i,j)$ is in **violation-tier** $k$ if $\alpha/2^k < v_{ij} \leq \alpha/2^{k-1}$. Observe that if $\alpha \geq \max_{i,j} w_{ij}$ then $v_{ij} \leq w_{ij}$. For any set of edges $S$, define $V(S) = \sum_{(i,j) \in S} v_{ij}$. Define $\tilde{v}_{ij} = \max\{v_{ij}/w_{ij}, 0\}$.*

The improved oracle is given in Algorithm 5.

---
**Algorithm 5** Improved Oracle for LP3.
---
1: Let $x_i = \frac{\alpha}{\sum_j u_j^t} u_i^t$.
2: Let $E_{violated,k} = \{(i,j) | (i,j) \text{ is in violation-tier } k\}$. for $k = 1, \cdots, K = \lceil \log_2 \frac{n}{\delta} \rceil$
3: Find a **maximal matching** $S_k$ in each $E_{violated,k}$.
4: Let $S = \cup_k S_k$ and $\Delta_V = V(S)$.
5: **if** $\Delta_V < \delta\alpha$ **then**
6:     For each $(i,j) \in S$, increase $x_i$ and $x_j$ by $2v_{ij}$.
7:     Further increase all $x_i$ by $\frac{\delta\alpha}{n}$. Return $\mathbf{x}$ and report failure.
8: **else**
9:     $S' \leftarrow \emptyset$.
10:     **repeat**
11:        Pick a edge $(i,j)$ from $S$ with largest $v_{ij}$ and add it to $S'$
12:        Eliminate all edges adjacent to $i$ or $j$ from $S$.
13:     **until** $S = \emptyset$
14:     Let $\Delta'_V = V(S')$.
15:     Return $y_{ij} = \tilde{v}_{ij}\alpha/\Delta'_V$ for $(i,j) \in S'$ and $y_{ij} = 0$ otherwise.
16: **end if**
---

**Lemma 12.** *In Algorithm 5, if $\Delta_V \geq \delta\alpha$ then we have a matching $\mathbf{y}$ such that $\sum_{(i,j) \in E} w_{ij}y_{ij} = \alpha$, and for all $i$ either $\mathrm{M}(i,\mathbf{y}) = -1$ or $\mathrm{M}(i,\mathbf{y}) \leq \frac{5}{\delta}\tilde{v}_{ij} - 1$ where $(i,j) \in S'$. As a consequence if $\Delta_V \geq \delta\alpha$ then Algorithm 5 returns an admissible solution with $\ell = 1$ and $\rho = \frac{5}{\delta}$.*

*Proof.* The proof is similar to the proof of Lemma 7, except that we will use violations (whereas the proof of Lemma 7 used the weights). Observe that since $y_{ij} = 0$ if $(i,j) \notin S'$ we have

$$\sum_{(i,j) \in E} w_{ij}y_{ij} = \sum_{(i,j) \in S'} w_{ij}y_{ij} = \sum_{(i,j) \in S'} w_{ij}\frac{\tilde{v}_{ij}\alpha}{\Delta'} = \frac{\alpha}{\Delta'_V} \sum_{(i,j) \in S'} v_{ij} = \alpha$$

Note $\sum_i x_i = \alpha = \sum_{(i,j) \in E} w_{ij}y_{ij}$ (observe $\alpha$ is not changed within an iteration).

Now suppose that $(i,j) \in S'$ is in tier $k$ and consider edges adjacent to $i$. All of them are in tier $k+1$ or higher and for each tier we have at most two edges (one adjacent to $i$ and one adjacent to $j$) because we pick a maximal matching for each tier. So the total violation of edges that are eliminated by $(i,j)$ is $\sum_{k'=1}^{\infty} \frac{2v_{ij}}{2^{k'-1}} \leq 4v_{ij}$. This shows that $V(S - S') \leq 4V(S')$ and therefore $V(S') \geq V(S)/5$. Hence $\alpha/\Delta'_V \leq 5\alpha/\Delta_V \leq 5/\delta$ (otherwise the oracle has failed).

Now $\mathrm{M}(i, \mathbf{y}) = \sum_{j:(i,j) \in E} y_{ij} - 1$. Therefore if $i$ is unmatched in $S'$ we have $M(i, \mathbf{y}) = -1$. Otherwise $M(i, \mathbf{y}) = y_{ij} - 1 = \tilde{v}_{ij}\alpha/\Delta'_V - 1 \leq \frac{5}{\delta}\tilde{v}_{ij} - 1$ where $(i,j) \in S'$. This proves the first part of the lemma.

For the second part observe that $\mathbf{c}^T \mathbf{y} = \sum_{(i,j) \in E} w_{ij} y_{ij} = \alpha$. Using Lemma 6 we have $\alpha \mathrm{M}(\mathcal{D}^t, \mathbf{y}) \leq \sum_{(i,j) \in E} y_{ij}(x_i^t + x_j^t - w_{ij})$. Now if $x_i^t + x_j^t - w_{ij} \geq 0$ then $y_{ij} = 0$ by construction. Therefore $\alpha \mathrm{M}(D^t, \mathbf{y}) \leq 0$ and so $\mathrm{M}(i, \mathbf{y}) \leq 0 \leq \delta$. Finally $0 \leq tv_{ij} \leq 1$ and therefore $-1 \leq \mathrm{M}(i, \mathbf{y}) \leq \frac{5}{\delta}$. The lemma follows. $\square$

**Lemma 13.** *If $\Delta_V < \delta\alpha$, then Algorithm 5 returns a feasible solution for LP3 with value at most $(1 + 5\delta)\alpha$.*

*Proof.* The proof follows similar arguments as in the proof of Lemma 8; except that we use violations in this proof instead of weights (as in the proof of Lemma 8). We consider the normalized weights $\mathbf{x}$ as a primal candidate for LP3. If the oracle fails, we augment $\mathbf{x}$ to obtain a feasible primal solution with a small increase.

Suppose that $(i,j)$ is in violation-tier $k$. Since $S_k$ was maximal, there exists an edge that is adjacent to either $i$ or $j$. So $x_i$ or $x_j$ is increased by at least $\alpha/2^{k-1}$ and the constraint is satisfied. If $(i,j)$ did not belong to any of the violation-tiers then its violation was less than $\delta\alpha/n$ and since $x_i, x_j$ are increased by $\delta\alpha/n$ this constraint is also satisfied.

For each edge $(i,j) \in S$, we increase the objective value by $2v_{ij}$ and $\sum_{(i,j) \in S} v_{ij} = \Delta_V < \delta\alpha$. Finally, we increase all $x_i$ by $\delta\alpha/n$ which increases the objective value by at most $\delta\alpha$. So our primal solution has value at most $(1 + 5\delta)\alpha$. $\square$

The next lemma is the central idea in this subsection. Consider running Algorithm 4, but with Algorithm 5 as the oracle instead of Algorithm 3. Based on Lemmas 12 and 13, and Theorem 11 we know that in $T = O(\frac{1}{\varepsilon^3} \log n)$ iterations we will find a $(1 - \epsilon)$ approximation of the maximum weighted matching. Surprisingly, we will now prove that even if we do not update the witness $\mathbf{y}$ for $\frac{1}{\delta}$ steps, the witness remains admissible!

**Lemma 14.** *Consider running Algorithm 4, but with Algorithm 5 as the oracle instead of Algorithm 3. If the dual witness $\mathbf{y}$ computed by the Algorithm 5) in iteration $t$ is admissible, then $\mathbf{y}$ remains admissible for all iterations $t + q$ where $q \leq 1/\delta$.*

*Proof.* Since $\mathbf{y}$ was admissible (in any iteration) $-\ell \leq \mathrm{M}(i, \mathbf{y}) \leq \rho$. Since $\mathbf{y}$ was computed in iteration $t$ we know from the proof of Lemma 12 that $\sum_{(i,j) \in E} w_{ij} y_{ij} = \alpha$ and $\mathrm{M}(i, \mathbf{y})/\rho \leq \tilde{v}_{ij}^t$. We use $\tilde{v}_{ij}^t$ to indicate that the fractional violation that was used to determine $y_{ij}$. Note that $w_{ij}(1 - \tilde{v}_{ij}) = x_i^t + x_j^t$. Also note $0 < \tilde{v}_{ij} \leq 1$ for a violated edge.

Note that even though we are not updating $\mathbf{y}$ across the iterations, $\mathbf{u}, \mathbf{x}$ are being updated (using the same $\mathbf{y}$ at every step) and we need to prove $\mathrm{M}(\mathcal{D}^{t+q}, \mathbf{y}) \leq \delta$ for every $t + q$ where $q \leq 1/\delta$.

Then by Lemma 6,
$$\mathrm{M}_q(\mathcal{D}^{t+q}, \mathbf{y}) = \frac{1}{\alpha} \sum_{(i,j) \in E} w_{ij} y_{ij} \frac{x_i^{t+q} + x_j^{t+q} - w_{ij}}{w_{ij}}$$

and since $\sum_{(i,j)\in E} w_{ij}y_{ij} = \alpha$, it is sufficient to show that $\frac{x_i^{t+q}+x_j^{t+q}-w_{ij}}{w_{ij}} \leq \delta$ for $y_{ij} \neq 0$. This means that we need to focus on the edges in the matching $S'$ only, since all other edges have $y_{ij} = 0$.

In the following $\mathbf{x}^{t+q}$ refers to the iterations of Algorithm 4 using the witness $\mathbf{y}$ at every step. Note $\epsilon = \delta/4\ell$ (since we will eventually set $\delta \ll 1$, and therefore $\epsilon \ll \frac{1}{2}$) and $\ell = 1$. Note that

$$x_i^{t+1} \leq \frac{(1+\delta/4)^{\tilde{v}_{ij}}}{(1-\delta/4)^{\delta/5}}x_i^t \quad \implies \quad x_i^{t+q} \leq x_i^t \frac{(1+\delta/4)^{q\tilde{v}_{ij}}}{(1-\delta/4)^{q\delta/5}}$$

The inequality on the left follows from $\rho = 5/\delta$. Note that we have the $1/(1-\delta/4)^{\delta/5}$ term because we decrease the weight of the unmatched vertices and then renormalize the total weight to $\alpha$ (we are inductively assuming that we did not report failure up to iteration $t+q-1$). The renormalization effectively increases the weight of the matched vertices by the same factor. For any $\delta > 0$ and $q \leq 1/\delta$, we have $(1+\delta/4)^q \leq (1+\delta/4)^{1/\delta} < e^{1/4} < 2$ and $1/(1-\delta/4)^{q\delta/5} \leq 1/(1-\delta/4)^{1/5} \leq 1+\delta$. Therefore,

$$x_i^{t+q} + x_j^{t+q} \leq (x_i^t + x_j^t)\frac{(1+\delta/4)^{q\tilde{v}_{ij}}}{(1-\delta/4)^{q\delta/5}} \leq (x_i^t + x_j^t)2^{\tilde{v}_{ij}^t}(1+\delta)$$

Since $2^{\tilde{v}_{ij}^t} \leq 1 + \tilde{v}_{ij}^t$ for $0 < \tilde{v}_{ij}^t \leq 1$ and $(x_i^t + x_j^t) = w_{ij}(1-\tilde{v}_{ij}^t)$ we have:

$$x_i^{t+q} + x_j^{t+q} \leq w_{ij}(1-\tilde{v}_{ij})(1+\tilde{v}_{ij})(1+\delta) \leq w_{ij}(1+\delta)$$

This implies that $\frac{x_i^{t+q}+x_j^{t+q}-w_{ij}}{w_{ij}} \leq \delta$ for all $y_{ij} \neq 0$ and therefore $\mathrm{M}_q(\mathcal{D}^{t+q}, \mathbf{y}) \leq \delta$. Now we can claim using using induction that $\mathbf{y}$ remains admissible for all $q \leq 1/\delta$ iterations (the inductive hypothesis was necessary to ensure that $\alpha$ did not change). The lemma follows. □

---

**Algorithm 6** Overall Algorithm for MWM.

1: In one pass, find a 6 approximate maximum matching using [10] and let $\alpha_0$ be the weight of the matching. Also ensure $\alpha_0 \geq w_{ij}$ for all $(i,j) \in E$.
2: $u_i^1 = 1$ for all $i \in [n]$ and $\alpha = 6\alpha_0$
3: **for** $t = 1$ to $T$ **do**
4:     Given $u_i^t$, run the oracle (Algorithm 5).
5:     If the oracle failed decrease $\alpha$ by factor $(1+\frac{\varepsilon}{3})$ and repeat line 4.
6:     Let $\mathrm{M}(i, \mathbf{y}^t) = \mathbf{A}_i\mathbf{y}^t - b_i$. ($\mathbf{y}$ is an admissible dual witness now)
7:     $u_i^{t+1} = \begin{cases} u_i^t(1+\epsilon)^{\mathrm{M}(i,\mathbf{y}^t)/5} & \text{if } \mathrm{M}(i, \mathbf{y}^t) \geq 0 \\ u_i^t(1-\epsilon)^{-\mathrm{M}(i,\mathbf{y}^t)/5} & \text{if } \mathrm{M}(i, \mathbf{y}^t) < 0 \end{cases}$
8: **end for**
9: Output $\frac{1}{T}\frac{1}{1+4\delta}\sum_t \mathbf{y}^t$.

---

**Theorem 15.** *Theorem 11 holds with $T = O(\frac{1}{\varepsilon^2}\log n)$ (and with $T+1$ passes) using Algorithm 6.*

*Proof.* We use Lemma 14 repeatedly. We can compute $\mathbf{y}^1$ (using Algorithm 4 and Algorithm 5 as oracle and use it for the next $\frac{1}{\delta}$ iterations. Observe that Lemma 14 shows that we cannot report failure within these $\frac{1}{\delta}$ iterations and $\alpha$ cannot change. Repeating the same argument we compute the witness only for every $\frac{1}{\delta}$ iterations. Observe that the overall algorithm simplifies to the description given in Algorithm 6.

Therefore we have $O(\frac{\delta}{\varepsilon^3}\log n) = O(\frac{1}{\varepsilon^2}\log n)$ actual computations of the dual witness, we have a $(1-\varepsilon)$ approximation, where $\delta = \varepsilon/30$. Computation of each $\mathbf{y}$ requires a pass. Note that we may need to repeat an iteration if the $\mathbf{y}$ was not admissible (as in Theorem 11) — but this only adds $O(\frac{1}{\varepsilon})$ iterations. The space requirement is $O(n(T+\log n))$ since we need to only remember the different $\mathbf{y}$ values we computed. □

# 5 Removing the Dependency on $n$

In this section, we present algorithms for MCM and MWM where the number of passes does not depend on the number of nodes. In each case we use the Algorithm 6 but we use a subgraph of the input graph and apply further analysis to bound the number of iterations $T$. Moreover, instead of starting from an initial state $u_i^1 = 1$ we will start the algorithm with different values of $u_i^1$.

We will also need to use the Corollary 2 instead of Theorem 1. Recall that the number of iterations of the multiplicative weight update framework is $O(\frac{1}{\delta^3}(\ln \max_i \frac{\Psi_i}{\Upsilon_i^1}))$ where $\Upsilon_i^t = u_i^t/(\sum_j u_j^t)$ and $\Psi_i$ is the upper bound of $\Upsilon_i^t$. Of these, $\frac{1}{\delta}$ iterations can be performed in a single pass. In what follows, we will reduce or bound the $(\ln \max_i \frac{\Psi_i}{\Upsilon_i^1})$ term. The key observation we will use in this regard is the following Lemma:

**Lemma 16.** *If $u_i^1 \geq w_{ij}$ for all $(i,j) \in E$ then during the execution of Algorithm 6, $x_i \leq 2u_i^1$.*

*Proof.* As the parameter $\alpha$ is decreased in Algorithm 6, (because a larger value of $\alpha$ ended up returning a primal feasible solution and so we are now decreasing $\alpha$) the value of $x_i$ decreases because **u** remains unchanged but $\alpha$ decreases. Thus it suffices to analyze the case when we do not change $\alpha$.

We first observe that if $x_i^t \geq w_{ij}$ for all $(i,j) \in E$ then vertex $i$ is not involved in any violations. Then no edge adjacent to $i$ can be chosen in **y** and we will have $M(i, \mathbf{y}) = \sum_{j:(i,j)\in E} y_{ij} - 1 = -1$. Then we will be setting $u_i^{t+1} = u_i^t(1-\epsilon)^{1/5}$ (based on the subroutine Algorithm 6). Moreover observe that $\sum_j u_j^{t+1} \geq \sum_j (1-\epsilon)^{1/5} u_j^t$, since for every $j$ we have $M(j, \mathbf{y}) \geq -1$. Therefore,

$$x_i^{t+1} = \frac{\alpha u_i^{t+1}}{\sum_j u_j^{t+1}} = \frac{\alpha u_i^t(1-\epsilon)^{1/5}}{\sum_j u_j^{t+1}} \leq \frac{\alpha u_i^t(1-\epsilon)^{1/5}}{\sum_j (1-\epsilon)^{1/5} u_j^t} = \frac{\alpha u_i^t}{\sum_j u_j^t} = x_i^t$$

Therefore $x_i$ can increase only if it is involved in some violation. But then $x_i < w_{ij} \leq u_i^1$. So the maximum value $x_i$ can achieve is when it is increased in a single step of Algorithm 6 to above $u_i^1$. The maximum value $u_i^{t+1}/u_i^t$ in a step of Algorithm 6 is $(1+\epsilon)^{M(i,\mathbf{y})/5} \leq (1+\epsilon)^{1/\delta}$. Note that $\epsilon \leq \frac{\delta}{4\ell}$ and therefore $(1+\epsilon)^{1/\delta} \leq e^{1/4}$. However we may also be decreasing $\sum_i u_i^t$; which can decrease by a factor $(1-\epsilon)^{1/5}$. Therefore $x_i$, which is the relative contribution of $u_i$ to $\sum_i u_i$ can increase at most by a factor of $e^{1/4}(1-\epsilon)^{-1/5}$ which is at most 2 for $\epsilon \leq \frac{1}{2}$. The lemma follows. □

In Algorithm 6 $\Upsilon_i^t = x_i^t/\alpha$. Note $\alpha \geq OPT/6$ after the first pass where $OPT$ is the maximum weighted matching. Therefore if $u_i^1 \geq w_{ij}$ for all $(i,j) \in E$ then using Lemma 16 $\Psi_i \leq \frac{12u_i^1}{OPT}$ and $\frac{\Psi_i}{\Upsilon_i^1} \leq \frac{12(\sum_j u_j^1)}{OPT} = O(\frac{(\sum_j u_j^1)}{OPT})$; where the last fact follows from the fact that $\sum_j u_j^1 \geq 2OPT$ since each $w_{ij}$ is less than $u_i^1, u_j^1$. Setting $\delta = \varepsilon/30$ we get a variant of Theorem 15 as follows:

**Theorem 17.** *If $w_{ij} \leq u_i^1$ for all edges $(i,j) \in E$ then for any $\varepsilon \leq \frac{1}{2}$ in $T$ passes where $T = O(\frac{1}{\varepsilon^2} \log \frac{(\sum_j u_j^1)}{OPT})$, and $O(n(T + \log n))$ space we can compute a $(1-\varepsilon)$ approximation for maximum weighted matching in bipartite graphs.*

## 5.1 The Simple Case of MCM

In this context $OPT$ denotes the size of the maximum cardinality matching in $G$. Consider the Algorithm 7 and the following lemma:

**Lemma 18.** *Let $OPT_S$ denote the size of maximum matching in the subgraph induced by the vertex set $S \subseteq V$, then (using the notation of Algorithm 7), we have $OPT - OPT_{S_{t+2}} \leq \frac{2}{3}(OPT - OPT_{S_t})$. This proof does not use bipartiteness.*

**Algorithm 7** A constant pass algorithm for maximum cardinality matching
1: Find a maximal matching and find a 2 approximation of $OPT$.
2: Let $S_0$ be the set of vertices that are matched.
3: **for** $t = 1$ **to** $\left\lceil \frac{\log(2/\varepsilon')}{\log(3/2)} \right\rceil$ **do**
4:     Find a maximal matching between $S_{t-1}$ and $V - S_{t-1}$. Let $T_t$ be the set of vertices in the maximal matching.
5:     $S_t = S_{t-1} \cup T_t$.
6: **end for**
7: Let $G'$ be a subgraph induced by $S_T$. This can be achieved by filtering the stream.
8: Run Algorithm 6 on $G'$

*Proof.* Fix an optimal solution in the original graph $G$ and an optimal solution in the subgraph induced by $S_t$. From the difference of two matchings, we can find $OPT - OPT_{S_t}$ vertex disjoint augmenting paths, say $\mathcal{P}$. We show that at least $\frac{1}{3}|\mathcal{P}|$ vertex disjoint augmenting paths are included in the graph induced by $S_{t+2}$.

Order the vertex disjoint augmenting paths $\mathcal{P}$ arbitrarily. Let $i' - i - Z - j - j'$ be the first path (where $Z$ is some sequence of vertices in $S_t$). Then $i', j' \notin S_t$ and $i' \neq j'$. In what follows we will show the condition $\mathcal{C}$ : we have an augmentation path $i'' - i - Z - j - j''$ available in $S_{t+2}$ for some $i'' \neq j'', i'', j'' \in V - S_t$ and $i'', j'' \in S_{t+2}$. (Note $\{i', j'\}$ can intersect $\{i'', j''\}$.)

If we prove this condition $\mathcal{C}$, then any augmentation path we find can remove at most two additional paths in $\mathcal{P}$ (since $i'', j''$ are now unavailable). Therefore we can find at least $\frac{1}{3}|\mathcal{P}|$ augmentations in $S_{t+2}$. This means that $OPT_{S_{t+2}} \geq OPT_{S_t} + \frac{1}{3}(OPT - OPT_{S_t})$ and therefore

$$OPT - OPT_{S_{t+2}} \leq OPT - \left(OPT_{S_t} + \frac{1}{3}(OPT - OPT_{S_t})\right)$$

Therefore if we prove this condition $\mathcal{C}$ the lemma follows. We now prove the condition $\mathcal{C}$. If both $i', j'$ were included in the matching in step $t+1$ or $t+2$ then the condition holds with $i'' = i'$ and $j'' = j'$ since we consider all edges in the induced subgraph. Therefore at least one of them, say $i'$, was not included in any of the two maximal matching in steps $t+1, t+2$. This means that $i$ was matched to some $\tilde{i}_1$ in step $t+1$ and some $\tilde{i}_2$ in step $t+2$ with $\tilde{i}_1 \neq \tilde{i}_2$. If $j' \neq \tilde{i}_1$ then $j$ or $j'$ was matched in step $t+1$ since the edge $(j, j')$ was available. If $j'$ was matched then the condition is satisfied with $i'' = \tilde{i}_1$ and $j'' = j'$. Otherwise $j$ was matched, say to $j''$, in step $t+1$ and $j'' \neq \tilde{i}_1$ since $i$ is matched to $\tilde{i}_1$ in the same matching. The condition is satisfied with $i'' = \tilde{i}_1$. Therefore the only remaining case is $j' = \tilde{i}_1$, but then the condition is satisfied with $i'' = \tilde{i}_2$ and $j'' = j'$. Therefore the lemma follows. □

If Lemma 18 is repeated as many times as in Algorithm 7, the difference between $OPT$ (the size of the optimal matching in $G$) and the optimal solution in the subgraph $G'$ is at most $\left(\frac{2}{3}\right)^{(\log \frac{2}{\varepsilon'})/(\log \frac{3}{2})} = 2^{-\log \frac{2}{\varepsilon'}} = \varepsilon'/2$ times $OPT$ (notice that $OPT - OPT_{S_0} \leq OPT$ since we started with a 2 approximation). Therefore $G'$ now contains a $(1 - \varepsilon'/2)$ approximation of the maximum matching in $G$. The size of each maximal matching is $O(|OPT|)$ and we repeat $O(\log \frac{1}{\varepsilon})$, the subgraph contains at most $O(|OPT| \log \frac{1}{\varepsilon})$ vertices. The number of passes to find the subgraph is $O(\log \frac{1}{\varepsilon})$ since we can find a maximal matching in one pass. Using Theorem 17 we have:

**Theorem 19.** *For any $\varepsilon \leq \frac{1}{2}$ Algorithm 7 provides a $(1 - \varepsilon)$ approximation for the maximum cardinality matching problem in bipartite graphs using $T = O(\frac{1}{\varepsilon^2} \log \log \frac{1}{\varepsilon})$ passes, and $O(n'(T + \log n'))$ space where $n' = \min\{n, |OPT| \log \frac{1}{\varepsilon}\}$. This implies a $\frac{2}{3}(1 - \varepsilon)$ result for general graphs* *using the integrality gap results of [14, 15]. The size of the matching can be computed in $O(n' \log n')$ space.*

Observe that to estimate the size we only need to remember $G'$ and the **u** both of which can be done using $O(n')$ space. The oracle (Algorithm 5) requires $O(n' \log n')$ space.

## 5.2 The Not So Simple Case of MWM

The weighted case is significantly more difficult than the unweighted case. The subgraph will now be expressed implicitly using the vertex weights $u_i$ as proxy. In the language of Linear Programming this means that, instead of staring from an uniform random sample of the constraints, we will start from a weighted sample. Let the maximum weighted matching be $\mathcal{M}$.

Before proceeding further, for the rest of this section we assume that the weights are discrete, i.e., $w_{ij} \in \{1, (1+\nu), \cdots, (1+\nu)^L\}$ where $L = O(\frac{1}{\nu} \log \frac{n}{\nu})$ to simplify the analysis for $\nu \leq 1/6$. This can be achieved in three steps and a single pass by: (i) using a single pass to find both a $\frac{1}{6}$ approximation of $w(\mathcal{M})$ using the algorithm of [10], and the maximum weight edge. Denote the larger of these two values by $w'$ (which is a lower bound on the weight of $\mathcal{M}$). (ii) deciding to ignore all edges of weight $\nu w'/n$ and (iii) deciding to multiply all edge weights by $n/(\nu w')$ and the performing the discretization by rounding down. Given any matching in this scaled setting, we have a matching in the original setting which is related by a simple scaling factor. Note that the discretization of the weights reduces the optimal solution by at most $(1-\nu)$ factor.

Given a discretized set of edges we run Algorithm 8. The Algorithm 8 that computes the weights of the vertices is similar to Algorithm 7, but is significantly non-trivial. Let $\mathcal{M}'$ denote the maximum weight matching in this new discretized setting and its weight be $w(\mathcal{M}')$. We ensure that:

C1: $\sum_i u_i^1 = \left(\frac{36}{\nu^3} \ln \frac{1}{\nu}\right) w(\mathcal{M}')$.

C2: Let $G' = (V, E')$ be a subgraph that consists of edges $(i,j)$ such that $w_{ij} \leq u_i^1, u_j^1$. Then, $G'$ contains a matching with weight at least $(1-3\nu)w(\mathcal{M}')$.

Hence, using Theorem 17 we obtain an $(1-\varepsilon)$-approximation of the maximum matching $\mathcal{M}''$ in $G'$ in $O(\frac{1}{\varepsilon^2} \log \frac{1}{\nu})$ passes. Then using C2 we would have a $(1-3\nu)(1-\varepsilon)$ approximation for the maximum weight matching $\mathcal{M}'$ in $G'$. This corresponds to a $(1-3\nu)(1-\varepsilon)(1-\nu)$ approximation for the maximum weight matching $\mathcal{M}$ in $G$ — the weights in the original graph are scaled differently, but the relationship is one-to-one. Setting $\nu = \frac{\varepsilon'}{8}$ and $\varepsilon = \varepsilon'/2$ we would get a $(1-\varepsilon')$ approximation of $\mathcal{M}$ (for all $\varepsilon' \leq \frac{1}{2}$). We now proceed to ensure C1 and C2.

**Lemma 20.** *(Condition C1.)* $\sum_i u_i^1 = \left(\frac{36}{\nu^3} \ln \frac{1}{\nu}\right) w(\mathcal{M}')$. *Bipartiteness is not used in this proof.*

*Proof.* For a vertex $i$ define $k(i)$ to be the maximum $k$ with $i \in S_k^q$. Therefore $u_i^1 = (1+\nu)^{k(i)}$. In what follows we will show a charging scheme where we charge $u_i^1$ to different edges in $\mathcal{M}'$. Consider the edge $e_i = (i,j)$ that caused the inclusion of $i$ to $S_{k(i)}^q$. Note that $|\mathcal{N}(i', k')| \leq q$ for all $i', k'$.

At least one of $i$ and $j$ must be matched in $\mathcal{M}'$, otherwise $\mathcal{M}'$ is not optimal. Moreover either $i$ or $j$ must have an edge adjacent to it in $\mathcal{M}'$ with weight at least $\frac{1}{2}(1+\nu)^{k(i)}$ (otherwise we can remove both those edges and add $e_i$ to increase the weight of $\mathcal{M}'$). Let the edge with the larger weight (between the two possible edges in $\mathcal{M}'$ adjacent to $i, j$) be $f(e_i)$. We charge $f(e_i)$ the value $u_i^1$. Note that $f(e_i), e_i$ are adjacent.

Now consider an edge $e = (i', j') \in \mathcal{M}'$ with $w_e = (1+\nu)^k$. This collects a charge for any vertex $i$ in level $k(i)$ such that $\frac{1}{2}(1+\nu)^{k(i)} \leq (1+\nu)^k = w_e$. In each such level $k(i)$, we can have

**Algorithm 8** A constant pass algorithm for maximum weighted matching.
1: $(i, j)$ is in **level** $k$ if $w_{ij} = (1 + \nu)^k$
2: **for each level $k = 0, 1, \cdots, L$ in parallel do**
3:     Find a maximal matching $E_k^0$.
4:     Let $C_k$ be the set of nodes matched in the maximal matching.
5:     Let $S_k^1 = C_k$
6:     **for $t = 1$ to $q = 8 \lceil \frac{1}{\nu^2} \ln \frac{1}{\nu} \rceil$ do**
7:         Find a maximal matching $E_k^t$ between $C_k$ and $V - S_k^t$.
8:         Let $T_k^t$ be the set of nodes matched in the maximal matching.
9:         $S_k^{t+1} = S_k^t \cup T_k^t$.
10:     **end for**
11:     Let $\mathcal{N}(i, k)$ denote the neighbors of vertex $i$ in $\cup_{t=0}^q E_k^t$.
12: **end for**
13: Let $u_i^1 = (1 + \nu)^k$ for the maximum $k$ with $i \in S_k^q$
14: Let $G' = (V, E')$ where $E' = \{(i, j) : w_{ij} \leq u_i^1, u_j^1\}$.
15: Run Algorithm 6 on $G'$ with initial weights $u_i^1$ and return its result.

$e = f(e_i)$ for at most $2q + 2$ different vertices $i$ since $e_i, e$ must be adjacent. If $e_i = (i', i)$ then there are at most $q + 1$ possibilities for $i$ (including $i'$). This is because either $i = i'$ or $i \in \mathcal{N}(i', k(i))$ and $|\mathcal{N}(i', k(i))| \leq q$. Counting the $e_i$ that arise from $j'$ as well, we know that $e = f(e_i)$ for at most $2q + 2$ vertices $i$. Therefore the charge on $e$ from vertices $i$ with the largest value of $k(i)$ is $2(q + 1)2w_e$. From the vertices that are in the immediately lower level, the charge is $\frac{4(q+1)w_e}{(1+\nu)}$. Summing over all the levels, the charge is

$$4(q+1)w_e \left(1 + \frac{1}{1+\nu} + \frac{1}{(1+\nu)^2} + \cdots \right) \leq \frac{4(q+1)}{\nu} w_e$$

Summing over all edges in $\mathcal{M}'$, since $\frac{4(q+1)}{\nu} \leq \frac{36}{\nu^3} \ln \frac{1}{\nu}$ we have the desired result (we use the fact that $q + 1 \leq \frac{9q}{8}$). $\square$

**Lemma 21.** *(Condition C2.) $G'$ contains a matching with weight at least $(1 - 3\nu)w(\mathcal{M}')$.*

*Proof.* We start with $\mathcal{M}'$ and modify it into a matching $\mathcal{F}$ so that $\mathcal{F}$ contains only the edges in $G'$. We charge the loss induced by the modification to the edges in $\mathcal{M}' \cup \mathcal{F}$. We first describe the modification procedure:

1. Initially $M = \mathcal{M}'$. $\mathcal{F} = \emptyset$. We will maintain $M \cup \mathcal{F}$ to be a matching.

2. Pick the edge in $M$ with the highest weight. Let this edge be $e = (i, j)$ with weight $w_e = (1 + \nu)^k$ in level $k$. Since $E_k^0$ was a maximal matching, either $i$ or $j$ is in $C_k$. Without loss of generality, let it be $i$. Thus $k(i) \geq k$. If $k(j) \geq k$ then both $u_i^1, u_j^1$ are at least $w_e$ and $e \in G'$. We add $(i, j)$ to $\mathcal{F}$ and remove $(i, j)$ from $M$. Therefore it suffices to consider $k(j) < k$ and $j \notin S_k^q$ then $i$ has at least $q$ neighbors in $S_k^q$ and $\mathcal{N}(i, k) = q$; since the edge $(i, j)$ was available for potential inclusion in the $q$ maximal matching.

    (a) If there is $i' \in \mathcal{N}(i, k)$ that is not matched in $M$ or $\mathcal{F}$. Then $k(i') \geq k$ and $(i, i') \in G'$. Add $(i, i')$ to $\mathcal{F}$ and remove $(i, j)$ from $M$.

    (b) Otherwise all $i' \in \mathcal{N}(i, k)$ is matched in $M$ or $\mathcal{F}$. If $i'$ is matched in $M$ denote its partner to be $\sigma(i', M)$. Otherwise $i'$ is matched in $\mathcal{F}$ and denote its partner as $\sigma(i', \mathcal{F})$.

i. If there exists at least $q/2$ vertices ($q$ is even) in $\mathcal{N}(i,k)$ which are matched in $F$, then delete $(i,j)$ from $M$ and every $(i', \sigma(i', \mathcal{F}))$ where $i' \in \mathcal{N}(i,k)$ a **red** charge of $\frac{2}{q}w_e$.

ii. If there exists $i' \in \mathcal{N}(i,k)$ which is matched in $M$ and its weight $w((i', \sigma(i', M))) < \nu w_e$, then we delete both $(i,j)$ and $(i', \sigma(i', M))$ from $M$ and add $(i', i)$ to $\mathcal{F}$. Note $(i', i) \in G'$. Then $(i', i)$ collects a **green** charge of $w((i', \sigma(i', M)))$.

iii. Otherwise, there exists at least $q/2$ vertices in $\mathcal{N}(i,k)$ which are matched in $M$, let this set be $Q_i$. Find the smallest weight edge in $M$ incident on a vertex in $Q_i$, let that vertex be $i_0$. Delete both $(i,j)$ and $(i_0, \sigma(i_0, M))$ from $M$ and add $(i, i_0)$ to $\mathcal{F}$. Each edge $(i', \sigma(i', M))$ where $i' \in \mathcal{N}(i,k)$ receives a **blue** charge of $\frac{2}{q}w_0$.

Observe that the sets $\mathcal{N}(i,k)$ are disjoint for a fixed $k$. This is because the matched vertices $S_k^t$ are ruled out from participating in step $t+1$ or later. Observe that the edges are added to $\mathcal{F}$ in non-increasing order of weight. Moreover, during the execution of the above procedure, at any point we have the invariant $\mathcal{I}$: that every edge in $\mathcal{F}$ has a weight at least at least as much as the heaviest weight edge in $M$.

The **red** charges are collected by edges in $\mathcal{F}$. Consider edge $e \in \mathcal{F}$ with $w_e = (1+\nu)^k$. Edge $e$ collects a **red** charge from edge $e'$ if $w_{e'} \leq w_e$. This is a consequence of the invariant $\mathcal{I}$. Moreover, for each $k' \leq k$ the edge $e$ can collect 2 such **red** charges for edges $e'$. This is because the two endpoints can be in $\mathcal{N}(i, k')$ for at most 2 different choices of $i$ (this is a consequence of $\mathcal{N}(i,k')$ being disjoint for a fixed $k$). The charge collected due to edges $e'$ in level $k'$ is $2\frac{2}{q}(1+\nu)^{k'}$. The the total **red** charge collected by $e$ can be bound by $\sum_{k'=0}^{k} 2\frac{2}{q}(1+\nu)^{k'} \leq \frac{4}{\nu q}(1+\nu)^k = \frac{4}{\nu q}w_e$. The overall **red** charge sums to $\frac{4}{\nu q}w(\mathcal{F})$.

The **green** charges are collected by edges in $\mathcal{F}$. Edge $e \in \mathcal{F}$ collects a **green** charge from edge $e'$ if $w_{e'} \leq \nu w_e$. Moreover, this charge is collected at most once. Therefore the total **green** charge is $\nu w(\mathcal{F})$.

The **blue** charges are collected by the edges in $\mathcal{M}'$. Consider edge $e \in \mathcal{M}'$ with $w_e = (1+\nu)^k$ which collect a **blue** charge when edge $e'$ was the heaviest weight edge in $M$ which was deleted from $M$ along with $e''$. Observe that since we were considering the edges in $M$ in decreasing order of weight, $w_{e'} \geq w_e$. Moreover $w_e \geq w_{e''}$, otherwise we would have deleted $e$ and charged $e''$ in that step. And finally, $w_{e''} \geq \nu w_{e'}$, otherwise we would be in the green case. Therefore we have $w_e \geq \nu w_{e'}$ and the edge $e$ is charged at most $\frac{2}{q}w_{e'} \leq \frac{2}{q}w_e$.

Let $w_{e'} = (1+\nu)^{k'}$ then $(1+\nu)^{k'} \geq (1+\nu)^k \geq \nu(1+\nu)^{k'}$ and we have $0 \leq k' - k \leq \frac{\ln \frac{1}{\nu}}{\ln(1+\nu)} \leq \frac{2}{\nu}\ln\frac{1}{\nu}$. The edge $e$ can collect 2 such **blue** charges for edges $e'$ in level $k'$ (again follows from $\mathcal{N}(i, k')$ being disjoint). The total **blue** charge on edge $e$ is at most $(\frac{2}{\nu}\ln\frac{1}{\nu})2\frac{2}{q}w_e = (\frac{8}{\nu q}\ln\frac{1}{\nu})w_e$. The overall **blue** charge sums to at most $\left(\frac{8}{\nu q}\ln\frac{1}{\nu}\right)w(\mathcal{M})$.

Observe that we maintained that $w(\mathcal{M}') = w(\mathcal{F}) + A$ where $A$ is the total charge. Putting the charges together, we have

$$w(\mathcal{M}') \leq w(\mathcal{F}) + \frac{4}{\nu q}w(\mathcal{F}) + \nu w(\mathcal{F}) + \frac{8}{\nu q}\ln\frac{1}{\nu}w(\mathcal{M}')$$

Using $q = 8\lceil \frac{1}{\nu^2} \ln \frac{1}{\nu} \rceil$ and rearranging we get

$$(1-\nu)w(\mathcal{M}') \leq \left(1 + \frac{\nu}{2\ln\frac{1}{\nu}} + \nu\right) w(\mathcal{F}) \leq (1+2\nu)w(\mathcal{F})$$

This translated to $w(\mathcal{F}) \geq \frac{1-\nu}{1+2\nu} w(\mathcal{M}') \geq (1-3\nu)w(\mathcal{M}')$ for all $\nu \leq \frac{1}{6}$ and the lemma follows. $\square$

It takes $q = O((\frac{1}{\varepsilon'})^2 \log \frac{1}{\varepsilon'})$ passes (for this setting of $\nu$) and $O(nL) = O(\frac{n\log n}{\varepsilon'})$ space to find the subgraph $G'$. Therefore (changing variables), we have

**Theorem 22.** *For any $\varepsilon \leq \frac{1}{2}$ in $T = O(\frac{1}{\varepsilon^2} \log \frac{1}{\varepsilon})$ passes, and $O(n(T + \frac{\log n}{\varepsilon}))$ space we can compute a $(1-\varepsilon)$ approximation for the maximum weighted matching in bipartite graphs. This translates to a $\frac{2}{3}(1-\varepsilon)$ approximation for general graphs using the integrality gap results of [14, 15]. The weight can be estimated using $O(\frac{n}{\varepsilon}\log n)$ space.*

Observe that to estimate the weight we need to compute and store the subgraph $G'$ which can be done using $O(\frac{n}{\varepsilon}\log n)$ space since we need to remember $O(n)$ vertices for each of the discretized weight levels. If we are only interested in the weight, the computation of the Algorithm 6 only needs $O(n\log n)$ space for the oracle and can remember $\mathbf{u}$ in space $O(n)$.

# 6 Extensions: the $b$-Matching Problem and the Maximum Matching Problem in General Graphs

In this section, we present algorithms for the $b$-matching problem and MWM in general graphs. Both algorithms are based on the idea from Section 4.1. In Section 6.1, we present algorithms for the capacitated $b$-matching problem in bipartite graphs (defined shortly). In Section 6.2 we discuss the uncapacitated $b$-matching problem in bipartite graphs. In Section 6.3, we present an approximation scheme for maximum weighted matching for general non-bipartite graphs.

## 6.1 The Maximum (Capacitated) bipartite $b$-Matching Problem

The stream is a sequence of tuples $\{(i,j,c_{ij},w_{ij})\}_{(i,j)\in E}$. We assume $c_{ij}, b_i$ are all integers for all $i, j$.

**Definition 5.** *In the (capacitated) $b$-matching problem, each vertex $i$ has demand $b_i$ and each edge $(i,j)$ has capacity $c_{ij}$ and weight $w_{ij}$. A multiset of edges is a $b$-matching if the multiplicity of each edge $(i,j)$ is at most $c_{ij}$ and $i$ is the endpoint of at most $b_i$ edges (counting the multiplicity of the edges) in the set. The maximum (capacitated) $b$-matching problem is to find a $b$-matching that maximizes the total weight of edges (again, accounting the multiplicity of the edges).*

We refer to $b_i$ as the capacity of a vertex $i$ and let $B = \sum_i b_i$ be the total capacity of all vertices. We assume that we have $\tilde{O}(B)$ space since the solution can have $O(B)$ edges in it. LP5 and LP6 are the primal and dual linear programs with integrality gap one [31]. Algorithm 9 is the oracle for LP5.

$$\begin{array}{ll}
\max & \sum_{(i,j)\in E} w_{ij} y_{ij} \\
\text{s.t} & \frac{1}{b_i}\sum_{(i,j)\in E} y_{ij} \leq 1 \quad \forall i \\
& \frac{1}{c_{ij}} y_{ij} \leq 1 \quad \forall (i,j) \\
& y_{ij} \geq 0 \quad \forall (i,j) \in E
\end{array} \quad \text{(LP5)} \qquad
\begin{array}{ll}
\min & \sum_i x_i + \sum_{(i,j)\in E} z_{ij} \\
\text{s.t} & \frac{x_i}{b_i} + \frac{x_j}{b_j} + \frac{z_{ij}}{c_{ij}} \geq w_{ij} \quad \forall (i,j) \in E \\
& x_i \geq 0 \quad \forall i
\end{array} \quad \text{(LP6)}$$

**Algorithm 9** Oracle for LP6.

1: Let $x_i = \frac{\alpha}{\sum_j u_j + \sum_{(i',j') \in E} u_{i'j'}} u_i$ and let $z_{ij} = \frac{\alpha}{\sum_j u_j + \sum_{(i',j') \in E} u_{i'j'}} u_{ij}$.
2: Let $E_{violated,k} = \{(i,j) | x_i/b_i + x_j/b_j + z_{ij}/c_{ij} < w_{ij}, \alpha/2^k < w_{ij} \leq \alpha/2^{k-1}\}$.
3: Find a maximal b-matching $S_k$ in $E_{violated,k}$ for each $k = 1, \cdots, \lceil \log(n/\delta) \rceil$.
4: Let $S = \cup_k S_k, \Delta = w(S)$.
5: Let $d_{ij}$ be the multiplicity of $(i,j)$ in $S$.
6: **if** $\Delta < \delta\alpha$ **then**
7:    For each $(i,j) \in S$, increase $x_i$ and $x_j$ by $2d_{ij}w_{ij}$ and $z_{ij}$ by $d_{ij}w_{ij}$.
8:    Further increase all $x_i$ by $\delta\alpha/n$. Return **x** and report failure.
9: **else**
10:    **repeat**
11:      Pick the heaviest edge $(i,j)$ from $S$.
12:      Add $(i,j)$ to $S'$.
13:      Suppose that $(i,j) \in S_k$.
14:      **for** $k' = k+1, k+2, \cdots$ **do**
15:         Reduce multiplicities of edges adjacent to $i$ and $j$ from $S_{k'}$ by $d_{ij}$ in total.
16:      **end for**
17:      Remove $(i,j)$ from $S$.
18:    **until** $S = \emptyset$
19:    Let $d'_{ij}$ be the multiplicity of $(i,j)$ in $S'$.
20:    Return $y_{ij} = \alpha d_{ij}/w(S')$ for $(i,j) \in S'$ and $y_{ij} = 0$ otherwise.
21: **end if**

**Computing $z_{ij}$:** The computation of $z_{ij}$ is also not trivial since we cannot store all values of $u_{ij}$. In one pass, we can count the number of edges and therefore, we know the number of constraints in LP5. Observe that $u_{ij}$ values are identical for all edges $(i,j)$ that have never been selected for $S'$. So if we remember $u_{ij}$ values only for edges that have been in $S'$, we can compute $z_{ij}$ for all edges.

**Lemma 23.** *If $\Delta \geq \delta\alpha$, Algorithm 9 returns an admissible solution **y** with $\ell = 1$ and $\rho = 5/\delta$.*

*Proof.* We first observe that $\sum_{(i,j) \in E} w_{ij}y_{ij} = \sum_{(i,j) \in S'} w_{ij}y_{ij}$ and

$$\sum_{(i,j) \in S'} w_{ij}y_{ij} = \sum_{(i,j) \in S'} \frac{\alpha w_{ij}d_{ij}}{w(S')} = \frac{\alpha}{w(S')} \sum_{(i,j) \in S'} w_{ij}d_{ij} = \frac{\alpha}{w(S')} w(S') = \alpha$$

Now observe that (dropping the superscript $t$ for this equation)

$$\begin{aligned}
\alpha M(\mathcal{D}^t, \mathbf{y}) &= \sum_i x_i \left( \frac{1}{b_i} \sum_{j:(i,j) \in E} y_{ij} - 1 \right) + \sum_{(i,j) \in E} z_{ij} \left( \frac{1}{c_{ij}} y_{ij} - 1 \right) \\
&= \sum_{(i,j) \in E} y_{ij} \left( \frac{x_i}{b_i} + \frac{x_j}{b_j} + \frac{z_{ij}}{c_{ij}} \right) - \left( \sum_i x_i + \sum_{(i,j) \in E} z_{ij} \right) \\
&= \sum_{(i,j) \in E} y_{ij} \left( \frac{x_i}{b_i} + \frac{x_j}{b_j} + \frac{z_{ij}}{c_{ij}} \right) - \alpha \quad \text{(By normalization.)} \\
&= \sum_{(i,j) \in E} y_{ij} \left( \frac{x_i}{b_i} + \frac{x_j}{b_j} + \frac{z_{ij}}{c_{ij}} \right) - \sum_{(i,j) \in E} w_{ij}y_{ij} = \sum_{(i,j) \in E} y_{ij} \left( \frac{x_i}{b_i} + \frac{x_j}{b_j} + \frac{z_{ij}}{c_{ij}} - w_{ij} \right)
\end{aligned}$$

Now if $\left(\frac{x_i}{b_i} + \frac{x_j}{b_j} + \frac{z_{ij}}{c_{ij}} - w_{ij}\right) \geq 0$ then $y_{ij} = 0$. Therefore $\alpha M(\mathcal{D}^t, \mathbf{y}) \leq 0 \leq \delta$.

Observe that $-1 \leq M(i, \mathbf{y}) \leq \frac{\alpha}{w(S')}$ since $\sum_j d_{ij} \leq b_i$. Likewise $-1 \leq M(ij, \mathbf{y}) \leq \frac{\alpha}{w(S')}$ since $d_{ij} \leq c_{ij}$.

Finally observe that for each edge $(i,j) \in S'$ we eliminate at most $2d_{ij}$ elements from each higher tier $k'$ (which means weight is lower). Therefore the total elimination is at most $4d_{ij}w_{ij}$ (same argument as in Lemma 7) which means $w(S - S') \leq 4w(S')$. Therefore $w(S) \leq 5w(S')$ and $\alpha/w(S') = 5\alpha/\Delta \leq 5/\rho$. The lemma follows. □

**Lemma 24.** *If $\Delta < \delta\alpha$, a feasible solution for LP6 with value at most $(1 + 6\delta)\alpha$ is returned.*

*Proof.* Suppose the edge $(i,j)$ was considered in one of the tiers. For each $(i,j) \in E$ with multiplicity $d_{ij}$ ($d_{ij} = 0$ for $(i,j) \notin S$), the algorithm does not give a higher value of $d_{ij}$ to $(i,j)$ because $(i,j), i$ or $j$ did not allow so in $S_k$. If $(i,j)$ was the problem, that is $d_{ij} = c_{ij}$, we increase $z_{ij}$ by $c_{ij}w_{ij}$ and the primal constraint for $(i,j)$ is satisfied. Otherwise one of the vertices $i, j$ had $b_i$ adjacent edges. Suppose it is $i$. Then we increase $x_i$ by at least $b_{ij}w_{ij}$ because all the edges adjacent to $i$ contribute twice their weight (times the respective multiplicity) and these edges are from tier from $k$ or lower (which means weight is at least half of $w_{ij}$). So the primal constraint is satisfied for $(i,j)$.

If $(i,j)$ was not considered in one of the tiers then $w_{ij} \leq \delta\alpha/n$. But then increasing $x_i$ by $\delta\alpha/n$ makes this constraint satisfied as well. Therefore, the returned solution is feasible. For each $(i,j) \in S$, we increase the objective value by $5d_{ij}w_{ij} = 5w(S) \leq 5\delta\alpha$. In addition we increase each $x_i$ by $\delta\alpha/n$ and the total increase is $6\delta\alpha$. □

We can now apply Theorem 1, using $\varepsilon/36$ and the space saving idea of Section 4.1) of slowly lowering the guess of $\alpha$. Note $|S'| = O(B)$ and $|S| = O(B \log n)$. In this case we do not have an easy $O(1)$ approximation. However we can easily guess a factor $n$ approximation and run the algorithm for $O(\frac{1}{\varepsilon} \log n)$ guesses of the optimum solution. Observe that the number of iterations is only additive in the number of guesses of the optimum solution (see Section 4.1). Since we only need to provision for a single copy of the oracle, we have:

**Theorem 25.** *For any $\varepsilon \leq \frac{1}{2}$, in $T = O(\frac{1}{\varepsilon^3} \log n)$ passes and $O(BT)$ space, Algorithm 9 and Algorithm 1 together provide a $(1 - \varepsilon)$ approximation for the optimum b-matching problem.*

Note that we only find a fractional solution in this case. Also note that the actual space requirement depends on $c_{ij}$ as well as $b_i$. As the minimum value of $c_{ij}$ increases, $|S|$ decreases and therefore we need less space. An extreme case of this is the uncapacitated problem, which we discuss next.

## 6.2 The Maximum Uncapacitated b-Matching Problem

In this section, we present two algorithms for the maximum uncapacitated b-matching problem: a constant pass algorithm that requires space that depends on $B$ and a near linear space algorithm that requires the number of passes that depends on $\log n$. The uncapacitated b-matching problem is a special case of the capacitated b-matching problem where all the capacities are infinite. LP7 and LP8 are the primal and dual LPs for the uncapacitated b-matching problem.

$$\begin{array}{ll} \max & \sum_{(i,j) \in E} w_{ij} y_{ij} \\ \text{s.t} & \frac{1}{b_i} \sum_{(i,j) \in E} y_{ij} \leq 1 \quad \forall i \\ & y_{ij} \geq 0 \quad \forall (i,j) \in E \end{array} \quad \text{(LP7)} \qquad \begin{array}{ll} \min & \sum_i x_i \\ \text{s.t} & \frac{x_i}{b_i} + \frac{x_j}{b_j} \geq w_{ij} \quad \forall (i,j) \in E \\ & x_i \geq 0 \quad \forall i \end{array} \quad \text{(LP8)}$$

$O(\frac{1}{\varepsilon^2} \log \frac{1}{\varepsilon})$**-pass** $O(B(\frac{1}{\varepsilon^2} \log \frac{1}{\varepsilon} + \frac{\log n}{\varepsilon}))$**-space algorithm:** The uncapacitated $b$-matching problem reduces to the maximum matching problem with $O(B)$ vertices [32, 30]. Each vertex $i$ with duplicity $b_i$ becomes $b_i$ vertices $i_1, i_2, \cdots, i_{b_i}$. Each edge $(i,j)$ becomes $b_i b_j$ edges $(i_1, j_1), (i_1, j_2), \cdots, (i_{b_i}, j_{b_j})$. Let the resulting graph be $G'$. It is easy to see that any $b$-matching in $G$ corresponds to a matching of the same weight in $G'$ and the converse also holds. Using Algorithm 8 for $G'$, we obtain the following result. This transformation preserves bipartiteness.

**Corollary 26** (Theorem 22). *For any $\varepsilon \leq \frac{1}{2}$ in $T = O(\frac{1}{\varepsilon^2} \log \frac{1}{\varepsilon})$ passes, and $O(B(T + \frac{\log n}{\varepsilon}))$ space we can compute a $(1-\varepsilon)$ approximation for the maximum weighted uncapacitated $b$-matching in bipartite graphs.*

$O(\frac{1}{\varepsilon^3} \log n)$**-pass** $\tilde{O}(\frac{n}{\varepsilon^3})$**-space algorithm:** Since the uncapacitated problem is a special case of the capacitated $b$-matching problem, we can apply Algorithm 9. However, the uncapacitated problem differs from the capacitated $b$-matching problem in that we can always find a maximal $b$-matching with at most $n-1$ (distinct) edges in the former. If we have a solution where $\sum_{(i,j) \in E} y_{ij} = b_i$ we denote the vertex $i$ to be saturated. Suppose that we are given an edge $(i,j)$ and neither of the vertices $i,j$ are saturated. We can saturate $i$ or $j$ by increasing $y_{ij}$. This process saturates one vertex while increases the number of edges in the $b$-matching. As a consequence it gives a maximal $b$-matching of at most $n-1$ edges which leads to the following corollary.

**Corollary 27** (Theorem 25). *For any $\varepsilon \leq \frac{1}{2}$ in $T = O(\frac{1}{\varepsilon^3} \log n)$ passes, and $O(nT)$ space we can compute a $(1-\varepsilon)$ approximation for the maximum weighted uncapacitated $b$-matching in bipartite graphs.*

## 6.3 The Maximum Weight Matching Problem for General Graphs

Algorithms in Sections 5 achieve $(\frac{2}{3} - \epsilon)$-approximations because the integrality gap of LP4 is $\frac{2}{3}$ for general graphs [14, 15]. With additional constraints, we can write a linear program for MWM in general graphs with integrality gap one. LP9 and LP10 are the primal and dual LP for general graphs. In addition to the constraints in MWM, we have a constraint for each odd subset $U$ of $V$. The polytope determined by constraints in LP9 is the convex hull of all matchings [6, 30].

$$
\begin{array}{ll}
\max & \sum_{(i,j) \in E} w_{ij} y_{ij} \\
\text{s.t} & \sum_{(i,j) \in E} y_{ij} \leq 1 \quad \forall i \\
& \sum_{i,j \in U} y_{ij} \leq \lfloor |U|/2 \rfloor \quad \forall U \\
& y_{ij} \geq 0 \quad \forall (i,j) \in E
\end{array}
\quad \text{(LP9)}
\qquad
\begin{array}{ll}
\min & \sum_i x_i + \sum_U z_U \\
\text{s.t} & x_i + x_j + \sum_{i,j \in U} \frac{1}{\lfloor |U|/2 \rfloor} z_U \geq w_{ij} \quad \forall (i,j) \in E \\
& x_i \geq 0 \quad \forall i
\end{array}
\quad \text{(LP10)}
$$

The violation and weight that correspond to the odd-set constraints are

$$
M(U, y^t) = \left( \frac{1}{\lfloor |U|/2 \rfloor} \sum_{i,j \in U} y_{ij}^t \right) - 1
$$

$$
u_U^t = \sum_{M(U,y^t) \geq 0} (1+\epsilon)^{M(U,y^t)/\rho} \cdot \sum_{M(U,y^t) < 0} (1-\epsilon)^{M(U,y^t)/\rho}
$$

Since there are exponentially many odd sets $U$, we cannot store all weights $u_U^t$. Instead, we remember non-zero values of $\mathbf{y}^t$ for all $t$. Then, we can recompute the values of $u_U^t$ given $U$ without reading data stream.

There is another problem due to the number of constraints. Since the number of constraints is exponential in $n$, the number of iterations is linear in $n$ where we want the number of iterations is polynomial in $\log n$ and $\frac{1}{\varepsilon}$. To reduce the number of constraints, we simply ignore all constraints corresponding to $U$ with $|U| > \frac{1}{\delta}$. Then, the number of constraints is $O(n^{\delta^{-1}})$ and therefore, the number of iterations is $O(\frac{1}{\delta^4} \log n)$. From a feasible solution for the modified formulation, we obtain a feasible solution for LP9 by scaling $y_{ij}$. If constraints corresponding to vertices are satisfied, $\sum_{i,j \in U} y_{ij} \leq (1+\delta)\lfloor |U|/2 \rfloor$ for $|U| > \frac{1}{\delta}$. Therefore, if we scale all $y_{ij}$ by factor $\frac{1}{1+\delta}$, we satisfy all the constraints in LP9.

---

**Algorithm 10** Oracle for LP10.

---

1: Keep $\mathbf{y}^{t'}$ for all $t' = 1, 2, \cdots, t-1$.
2: $W \leftarrow \sum_i u_i^t$
3: **for each** $U \subset V, |U|$ is odd **do**
4:  Compute $u_U^t$ where

$$u_U^t = (1+\frac{\delta}{4})^{\left(\sum_{M(U,y^{t'}) \geq 0} \delta M(U, y^{t'})\right)} \cdot (1-\frac{\delta}{4})^{\left(\sum_{M(U,y^{t'}) < 0} \delta M(U, y^{t'})\right)}$$

5:  $W \leftarrow W + u_U^t$.
6: **end for**
7: Let $x_i = \frac{\alpha}{W} u_i^t$, $z_U = \frac{\alpha}{W} u_U^t$. Instead of storing all $z_U$, $z_U$ is recomputed for each $(i,j) \in U$.
8: Let $E_{violated,k} = \{(i,j) | x_i + x_j + \sum_{i,j \in U} \frac{1}{\lfloor |U|/2 \rfloor} z_U < w_{ij}, \frac{\alpha}{2^k} < w_{ij} \leq \frac{\alpha}{2^{k-1}}\}$.
9: Find a maximal matching $S_k$ in $E_{violated,k}$ for each $k = 1, \cdots, \lceil \log(n/\delta) \rceil$.
10: Let $S = \cup_k S_k, \Delta = w(S)$.
11: **if** $\Delta < \delta\alpha$ **then**
12:  For each $(i,j) \in S$, increase $x_i$ and $x_j$ by $2w_{ij}$.
13:  Further increase $x_i$ by $\delta\alpha/n$. Return $\mathbf{x}$.
14: **else**
15:  **repeat**
16:   Pick a heaviest edge $(i,j)$ from $S$ and add it to $S'$
17:   Eliminate all edges adjacent to $i$ or $j$ from $S$.
18:  **until** $S = \emptyset$
19:  Return $y_{ij} = \alpha/w(S')$ for $(i,j) \in S'$ and $y_{ij} = 0$ otherwise.
20: **end if**

---

Algorithm 10 is the oracle for LP10. It is similar to Algorithm 3. The proof of correctness is almost identical to Section 3. We also use the ideas in Section 4.1 which hold in this case as well. One drawback of this algorithm is its running time. For each $(i,j)$, we have to enumerate all $U$ that contain $(i,j)$. There are $n^{O(\varepsilon^{-1})}$ subsets that contain $(i,j)$. For each $U$, it takes $\tilde{O}(\frac{1}{\varepsilon^5})$ time to compute the weight because there are at most $O(\frac{1}{\varepsilon})$ edges for each $t'$. Therefore, it takes $\tilde{O}(\frac{n^{O(\varepsilon^{-1})}m}{\varepsilon^5})$ time per pass. We obtain the following theorem:

**Theorem 28.** *In $T = O(\frac{1}{\varepsilon^4} \log n)$ passes and $\tilde{O}(\frac{n^{O(\varepsilon^{-1})}m}{\varepsilon^9})$ time, we can compute a $(1-\varepsilon)$ approximation for maximum weighted matching in general graphs. The algorithm uses $O(nT)$ space.*